\documentclass[12pt]{article}
\usepackage{epsfig,graphicx}
\usepackage{amsmath,amsfonts,amssymb} 
\usepackage{color}
\usepackage{hyperref}
\usepackage{cite}
\usepackage{slashed}
\usepackage{graphicx}
\usepackage{pgfplots}
\pgfplotsset{compat=newest}
\usepackage{color}
\usepackage{graphics,cite,amssymb,float,ifpdf}
\usepackage{amsmath, mathbbol}
\usepackage{amsfonts}
\usepackage{xr}

\definecolor{vmlgreen}{rgb}{0.0, 0.5, 0.0}

\newcommand{\unit}{\leavevmode\hbox{\small1\kern-3.6pt\normalsize1}}

\newcommand{\squared}[1]{\left| #1 \right|^2}
 
\newcommand{\gsim}{\raisebox{-0.13cm}{~\shortstack{$>$ \\[-0.07cm] $\sim$}}~} 

\newcommand{\beq}{\begin{eqnarray}} 
\newcommand{\eeq}{\end{eqnarray}}

\allowdisplaybreaks

\begin{document}

\thispagestyle{empty}
\begin{flushright}
\rightline{IFT-UAM/CSIC-16-020}
\rightline{FTUAM-16-7}
  \vspace*{2.mm} \today
\end{flushright}

\begin{center}
  {\Large \textbf{Confronting dark matter with the diphoton excess from a parent resonance decay
  } }  
  
  \vspace{0.5cm}
  Valentina De Romeri$^{1,2}$, 
  Jong Soo Kim$^1$,
  V\'ictor Mart\'in-Lozano$^{1,2}$,\\
  Krzysztof Rolbiecki$^{1,3}$,
  Roberto Ruiz de Austri$^4$ \\[0.2cm] 
    
  {\small \textit{ 
      $^1$Instituto de F\'{\i}sica Te\'{o}rica UAM/CSIC, \\
      Universidad Aut\'onoma de Madrid, Cantoblanco, E-28049, Madrid, Spain\\
      $^2$Departamento de F\'{\i}sica Te\'{o}rica,\\ Universidad Aut\'{o}noma de Madrid, 28049
      Madrid, Spain \\
      $^3$Institute of Theoretical Physics, University of Warsaw, PL-02093 Warsaw, Poland\\[0pt] 
      $^4$Instituto de F\'{i}sica Corpuscular, IFIC/CSIC, Valencia, Spain\\
      }}
  
\vspace*{0.7cm}

\begin{abstract}
A diphoton excess with an invariant mass of about 750 GeV has been recently reported by both ATLAS and CMS experiments at LHC. While the simplest interpretation requires the resonant production of a 750 GeV (pseudo)scalar, here we consider an alternative setup, with an additional heavy parent particle which decays into a pair of 750 GeV resonances.  This  configuration improves the agreement between the 8 TeV and 13 TeV data. 
Moreover, we include a dark matter candidate in the form of a Majorana fermion which interacts through the 750 GeV portal. The invisible decays of the light resonance help to suppress additional decay channels into Standard Model particles in association with the diphoton signal.  We realise our hierarchical framework in the context of an effective theory, and we analyse the diphoton signal as well as the consistency with other LHC searches. We finally address the interplay of the LHC results with  the dark matter  phenomenology, namely  the compatibility with the relic density abundance and the indirect  detection bounds.

\end{abstract}

\end{center}

\newpage
	
\section{Introduction}
\label{sec:intro}
ATLAS and CMS have recently reported a modest excess in the search for Higgs-like resonances in the diphoton channel at an invariant mass around 750 GeV with a local significance of 3.6$\sigma$ and $2.6\sigma$, respectively \cite{ATLASdiphoton2015,CMSdiphoton2015}. ATLAS, with 3.2 fb$^{-1}$ of collected data, has found  an excess of 14 events in the signal region, whereas CMS had a somewhat lower integrated luminosity of 2.6 fb$^{-1}$ with a relatively mild excess consisting of 5 diphoton events. Including the look-elsewhere-effect the significances drop respectively to 2.6$\sigma$ and 1.2$\sigma$. Should the significance of this excess increase with more accumulated data, it would indicate the existence of New Physics beyond the Standard Model (SM). 
A simple explanation of the excess consists in the resonant production of a (pseudo)scalar with an invariant mass of 750 GeV and a relatively large branching ratio into the diphoton channel \cite{Landau:1948kw,Yang:1950rg}. A production cross section times diphoton branching ratio between 5 and 10 fb  fits the {\it observed} excess. However, the diphoton {\it signal} might be in slight tension with LHC Run-I data~\cite{Kim:2015ksf} since no significant excess was reported in the 8 TeV searches by both collaborations \cite{Aad:2014ioa,Aad:2015mna,Khachatryan:2015qba,CMS-PAS-EXO-12-045}. 

Here,  in order to ameliorate the tension between 8 and 13 TeV data, we consider a scenario with a heavy messenger resulting in a more involved event topology \cite{Kim:2015ron,Franceschini:2015kwy,Knapen:2015dap}. In this framework, we assume a heavy pseudoscalar parent resonance $\phi_2$ decaying into a pair of lighter 750 GeV pseudoscalar resonances $\phi_1$. No additional particles seem to accompany the diphoton signal, since the events in the sideband and in the signal region look very similar. Therefore, the decay modes of $\phi_1$ into {\it visible} particles other than photons must be suppressed.  The simplest solution is to assume that the lighter resonance mainly decays into an invisible particle $\psi$. Depending on the model assumptions, the largest observable final state could possibly  be $\gamma\gamma \psi\psi$. Moreover, if the relation between masses is approximately $m_{\phi_2}\approx 2m_{\phi_1}$, the lighter resonance would be produced at rest, resulting in little net missing momentum. 

The results of ATLAS favour a large width of the resonance, of about 45 GeV, with a local significance increasing up to 3.9$\sigma$ under this assumption. Instead, CMS slightly prefers scenarios with a narrow width. Should future results point to a large width, a large branching ratio into invisible particles would allow to accommodate this observation without invoking strongly interacting New Physics scenarios~\cite{Backovic:2015fnp}.

It is natural to identify the invisible particle $\psi$ with the dark matter (DM) (see for instance Ref.~\cite{DM750} for a non-exhaustive list of works on this topic). Among the plethora of DM candidates present in the literature, a weakly interacting massive particle (WIMP) produced via thermal freeze-out is one of the most appealing (see e.g.~\cite{Bertone:2004pz}). The 750 GeV resonance would then be identified as a portal to the WIMP DM sector similar to the well studied Higgs portal models \cite{Djouadi:2011aa}. 

In this work, we want to study the implications of the diphoton excess in a heavy parent resonance scenario on the DM phenomenology, assuming that the lighter 750~GeV resonance $\phi_1$ mediates the interactions of a Majorana spin-$\frac{1}{2}$ DM particle $\psi$. Although the interaction of (pseudo)scalars with the SM gauge bosons would typically require new heavy states, here we consider a model independent approach, where we do not specify precisely the heavy particle sector. Instead, we describe the interactions of both the two pseudoscalars and the DM via effective operators. We define a cutoff scale  $\Lambda_{\phi_i}$, where the heavy degrees of freedom are integrated out and in this way we assume the results here derived to be valid for any specific ultraviolet (UV) completion with the same degrees of freedom below $\Lambda_{\phi_i}$. We consider specific patterns for the effective couplings of this ``toy model" motivated by some realistic models. At this scope, we study two generic scenarios: one where the coupling of the lightest pseudoscalar $\phi_1$ to gluons is set to zero, and a second one where $\phi_1$ can couple to gluons as well as to the electroweak (EW) gauge bosons.
While we fit both scenarios to the diphoton excess, we also carefully check other LHC constraints including monojet and dijet searches and jets plus $E_T^{\mathrm{miss}}$ searches. Moreover, we investigate the DM phenomenology, taking into account cosmological and astroparticle constraints arising from the relic density abundance measured by the PLANCK satellite \cite{Ade:2013zuv} as well as from indirect detection (ID) searches with the Fermi-LAT satellite \cite{Ackermann:2015zua,Ackermann:2015lka}.

The paper is organised as follows. In the next Section, we present a simple model independent framework for the heavy parent resonance model. We fit the model parameters to the diphoton excess scrutinising the compatibility with LHC constraints in Section~\ref{sec:diphoton}. We then address the DM phenomenology in Section~\ref{sec:dm}, considering the constraints from cosmology and astroparticle physics. Finally, we  conclude with a brief summary in the last Section.

\section{Effective Lagrangian for the diphoton excess and the dark matter}\label{sec:model}
We consider a simple extension of the SM with the addition of two SM gauge singlet pseudoscalars $\phi_1$ and $\phi_2$. The kinetic and mass terms of both pseudoscalars are given by:
\begin{equation}\label{eq:kin}
\mathcal{L}_{\rm \phi}=\frac{1}{2}\partial_\mu\phi_1\partial^\mu\phi_1+\frac{1}{2}\partial_\mu\phi_2\partial^\mu\phi_2 -\frac{1}{2}m_{\phi_1}^2\phi_1^2-\frac{1}{2}m^2_{\phi_2}\phi_2^2\,,
\end{equation}
where $m_{\phi_1}$ and $m_{\phi_2}$ denote the masses of $\phi_1$ and $\phi_2$, respectively. We consider the following hierarchy: $2\times m_{\phi_1}\le m_{\phi_2}$.
The heavy resonance $\phi_2$ is coupled to the lighter resonance via a simple parity violating trilinear interaction term:
\begin{equation}\label{eq:trilinear}
\mathcal{L}_{\rm trilinear}=\lambda\phi_1\phi_1\phi_2\,.
\end{equation}
We assume that the couplings between both pseudoscalars and SM fermions via higher dimensional operators can be neglected. $\phi_1$ and  $\phi_2$ communicate with the SM sector via interactions with the SM gauge bosons parametrised by the following model independent effective couplings:
\begin{equation}\label{eq:gauge}
\mathcal{L}_{\rm  interactions}=\frac{c_3^{\phi_i}}{\Lambda_{\phi_i}}\epsilon^{\mu\nu\rho\sigma}G_{\mu\nu}^aG_{\rho\sigma}^a\phi_i+\frac{c_2^{\phi_i}}{\Lambda_{\phi_i}}\epsilon^{\mu\nu\rho\sigma}W^m_{\mu\nu}W^m_{\rho\sigma}\phi_i+\frac{c_1^{\phi_i}}{\Lambda_{\phi_i}}\epsilon^{\mu\nu\rho\sigma}B_{\mu\nu}B_{\rho\sigma}\phi_i\,,
\end{equation}
with $i=1,2$. Here, $c_j^{\phi_i}$, with $j=1,2,3$, are the effective couplings of the $\phi_i$ to the $SU(3)_C\times SU(2)_L\times U(1)_Y$ SM gauge bosons $G_\mu$, $W_\mu$ and $B_\mu$, respectively. 
$G_{\mu\nu}^a$, $W_{\mu\nu}^m$, $B_{\mu\nu}$ and $\Lambda_{\phi_i}$ correspond to the field strength tensors and the cut-off scale.\footnote{We always consider $\Lambda_{\phi_i}$ $\mathcal{O}(\mathrm{few\,\, TeV})$ in order to satisfy the effective field theory hypothesis.} $\epsilon_{\mu\nu\rho\sigma}$ is the totally antisymmetric tensor with $\epsilon_{0123}=+1$, $a=1,2,3$ and $m=1,2$ denotes $SU(3)_C$ and $SU(2)_L$ gauge indices, respectively. 
The prefactors $c_j^{\phi_i}$ are {\it a priori} free parameters but they can be explicitly calculated once the UV completion is known. For instance, the $c_j^{\phi_i}$ could be interpreted as anomaly induced couplings \cite{Wess:1971yu}. 

However, the goal of this work is not to discuss a specific model with UV completion, hence we do not give a thorough definition of the full particle spectrum beyond the SM. Nevertheless, the choice of the coefficients cannot be completely arbitrary either, hence we will later consider two different scenarios which can be motivated by some underlying UV physics.

We want to conclude this Section with a discussion of our final ingredient: the DM sector. There exists a large number of DM scenarios and in this work, we consider a hidden Majorana particle $\psi$ \footnote{The discussion for a Dirac DM candidate is straightforward.}.
 Its stability can be ensured by introducing a discrete $Z_2$ symmetry, i.e. only pairs of $\psi$'s couple to other particles,
\begin{equation}
\psi\rightarrow -\psi\,.
\end{equation}
 We assume that $\psi$ only couples to the lighter pseudoscalar via a Yukawa-type interaction with strength $g_\psi$:
\begin{equation}\label{eq:DM}
\mathcal{L}_{\rm DM}=i\bar{\psi}(\slashed{\partial}-m_\psi)\psi + ig_\psi\bar{\psi}\gamma_5\psi \phi_1\,.
\end{equation}
Thus, the relic density abundance of the DM candidate is governed by the $s$-channel exchange of $\phi_1$. 
It is difficult to accommodate a resonance with a large width assuming dominant couplings to {\it visible} particles, since strict limits exist on the SM decay modes of heavy resonances.
However, here we consider scenarios with $m_{\psi}<\frac{1}{2}m_{\phi_1}$ allowing for invisible decays of $\phi_1$. Depending on the size of $g_\psi$, the branching ratio into DM pairs can be sizeable. In principle, a very large invisible branching ratio allows for scenarios with large widths for $\phi_1$ -- $\Gamma(\phi_1)\sim \mathcal{O}(10)$ GeV -- as favoured by ATLAS diphoton data. 
\section{The diphoton signal and LHC constraints}\label{sec:diphoton}
\subsection{Heavy parent resonance}
\begin{figure}[!htb]
\begin{center}
\includegraphics[clip, trim=0.cm 17cm 0.cm 2cm,width=0.7\textwidth]{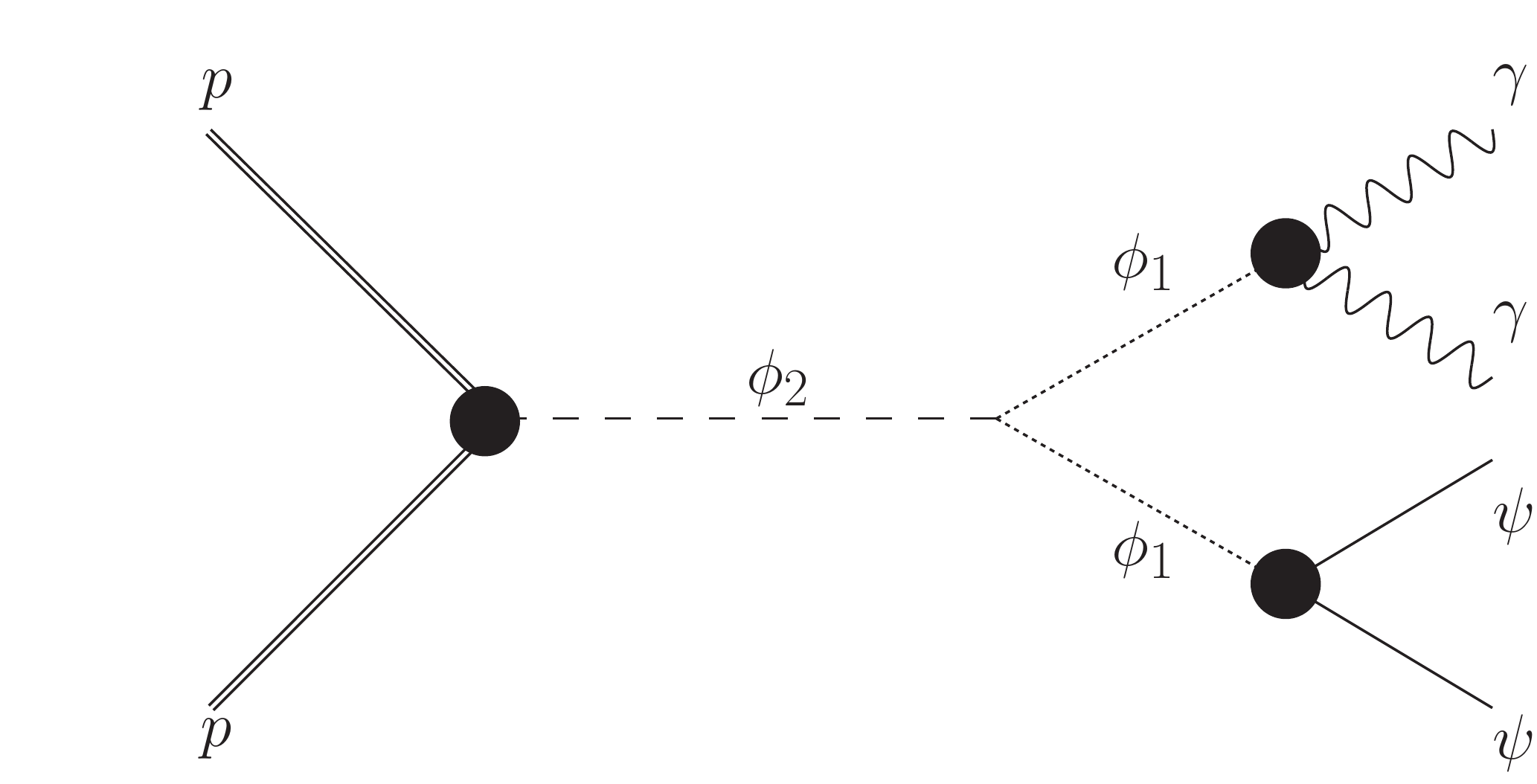}
\caption{The resonant production of $\phi_2$ followed by the decay to two 750 GeV $\phi_1$ pseudoscalars subsequently decaying into the diphoton and DM final state.}
\label{fig:process}
\end{center}
\end{figure}

In this Section, we discuss how to accommodate a diphoton signature from a 750~GeV resonance in our hierarchical framework. Our goal is to  
explain the excess as a result of the decay of a heavy parent resonance.
In this setup, we consider the production of a pseudoscalar resonance which decays into a pair of 750 GeV resonances $\phi_1$ subsequently decaying into SM gauge bosons as well as into DM. We expect the dominant diphoton signal from the following process, shown in Fig.~\ref{fig:process}:
\begin{equation}\label{eq:decaychain}
pp\rightarrow\phi_2\rightarrow\phi_1\phi_1\rightarrow\gamma\gamma+X\,,
\end{equation}
where $X$ denotes either $\psi\psi$ or SM gauge boson pairs. 

In order to achieve a dominant indirect production, the direct production of an $s$-channel resonance $\phi_1$ with an invariant mass of 750 GeV must be heavily suppressed. The typical cross section for the $\phi_1$ resonant production is given by\footnote{We consider dominant gluon fusion production. This applies to Scenario 2, where the coupling of $\phi_1$ to gluons is non-vanishing. We have numerically checked that the additional contribution from photon fusion is subdominant for our choices of the effective couplings.}:
\begin{equation}\label{eq:phi1}
\sigma (pp\rightarrow \phi_1\rightarrow\gamma\gamma)=\frac{1}{m_{\phi_1}s}\times C_{gg}\times \Gamma(\phi_1\rightarrow gg)\times \frac{\Gamma(\phi_1\to \gamma\gamma)}{\Gamma_{\phi_1}},
\end{equation}
where $s$ is the centre-of-mass energy and $C_{gg}=\frac{\pi^2}{8}\int_{m_{\phi_1}^2/s}^{1}\frac{dx}{x}g(x)g(\frac{m_{\phi_1}^2}{sx})$, with $g(x)$ the gluon distribution function. The numerical value of $C_{gg}$ for a mass of 750 GeV is $2137$ at $\sqrt{s}=13$ TeV using the gluon distribution function of Ref.~\cite{Martin:2009iq}. Analytical expressions for the partial decay widths $\Gamma(\phi_1\rightarrow gg)$ and $\Gamma(\phi_1\to \gamma\gamma)$ can be found in the Appendix, together with the other decay widths and the corresponding squared matrix elements for both pseudoscalars. Moreover, $\Gamma_{\phi_1}$  denotes the total decay width of $\phi_1$.  

\begin{figure}
\begin{center}
\includegraphics[width=.50\textwidth]{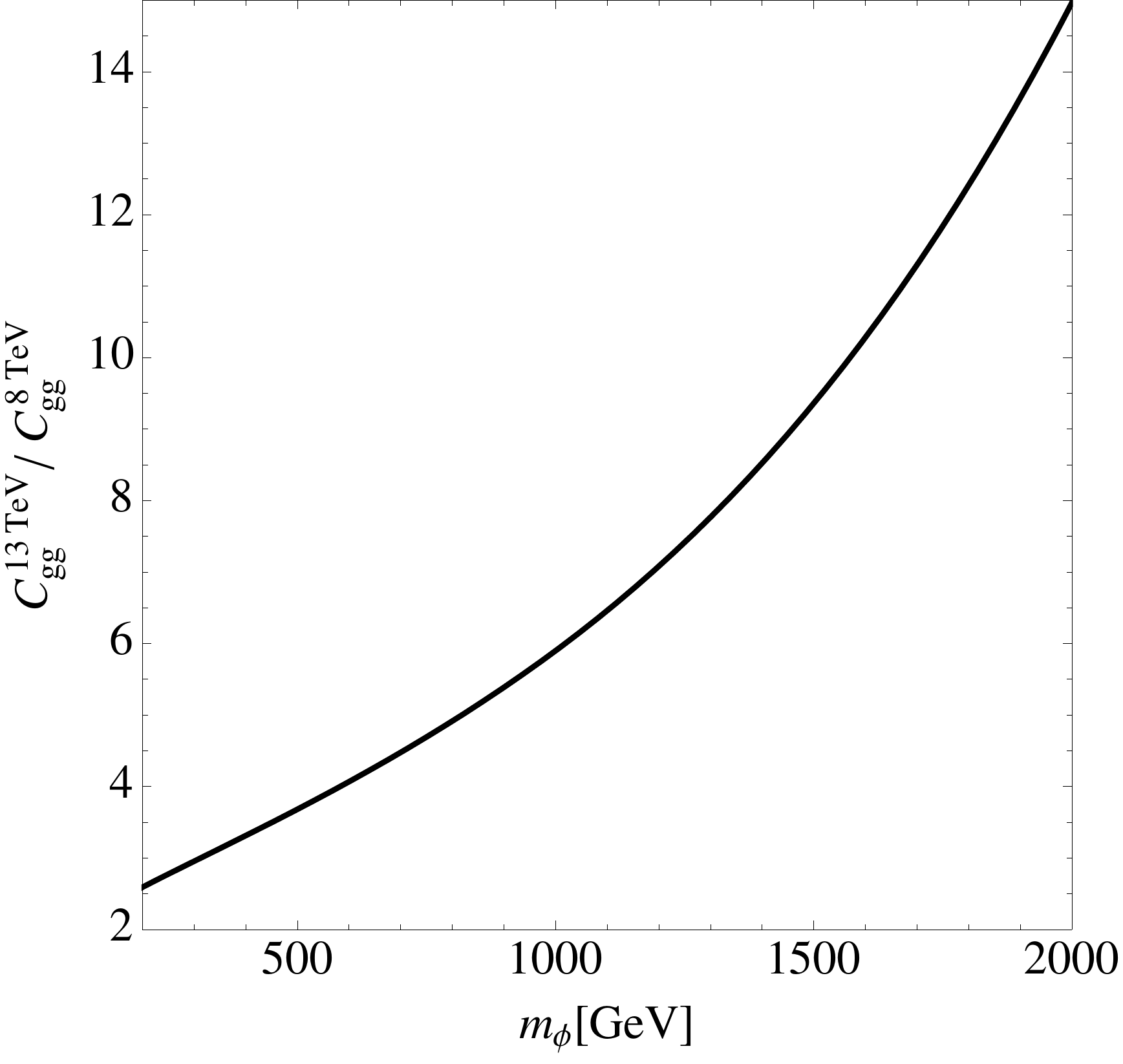}
\caption{Ratio of $C_{gg}$'s evaluated at $\sqrt{s}=$13~TeV and $\sqrt{s}=8$~TeV  as a function of the resonance mass. }
\label{fig:ratiocgg}
\end{center}
\end{figure}

The numerical evaluation of Eq.~(\ref{eq:phi1}) is straightforward. We show in Fig.~\ref{fig:ratiocgg} the ratio of the $C_{gg}$ evaluated at centre-of-mass energies of 13 TeV and 8 TeV as a function of the resonance mass. The cross section increases by a factor of $5$ for a 750~GeV resonance while rising the center-of-mass energy from 8 to 13 TeV. An even stronger increase in the production cross section can be gained for larger resonance masses and thus the effective accumulated data from Run-II can already be larger than Run-I for very heavy resonances. As a result, a diphoton excess originating from a parent resonance shows less tension between 8 and 13 TeV data as advertised in the Introduction.

\begin{figure}
\begin{center}
\includegraphics[width=.5\textwidth]{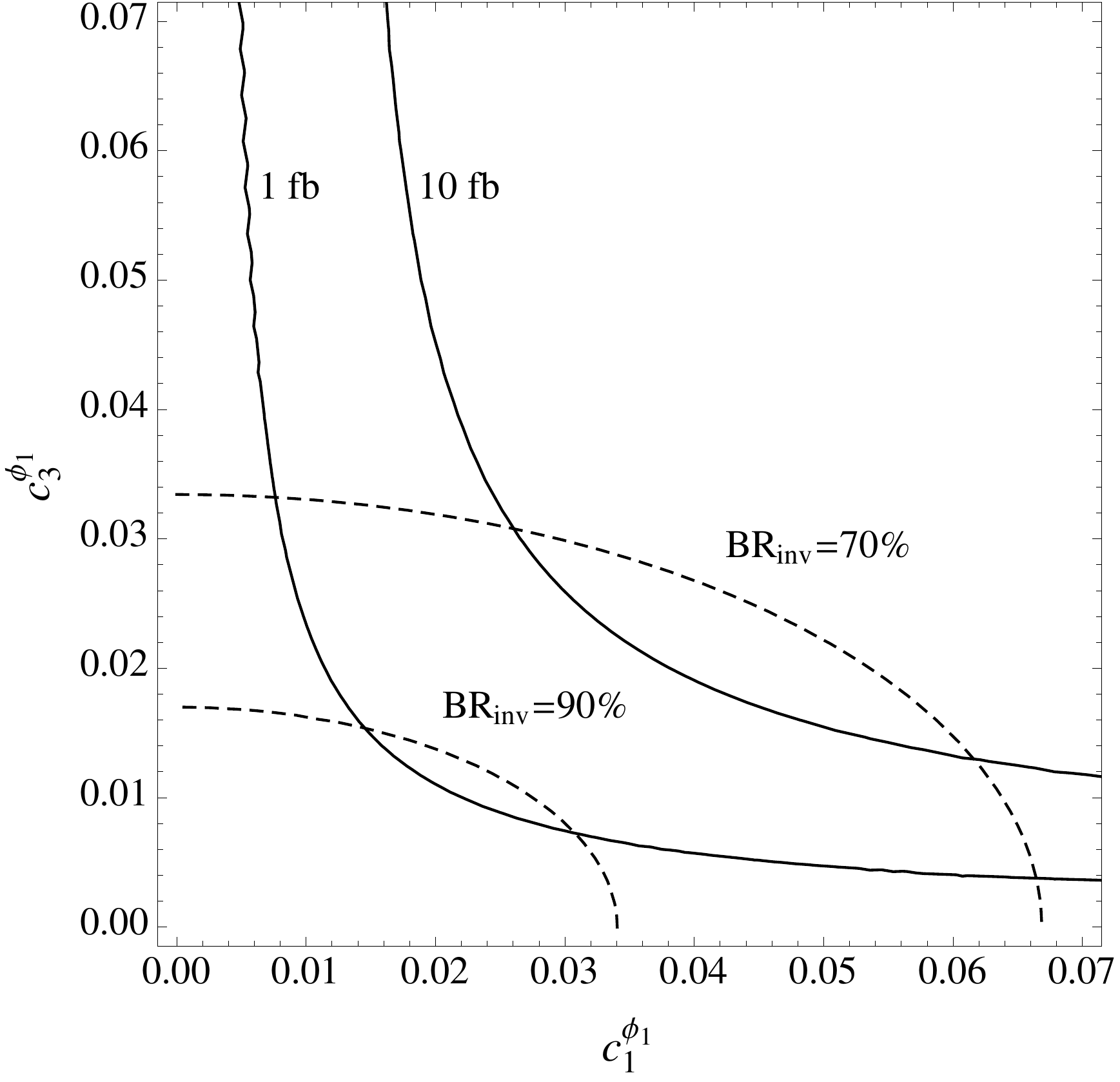}
\caption{The resonant production cross section of $\phi_1$ into diphoton final state (solid lines) in the ($c_3^{\phi_1}$,$c_1^{\phi_1}$) plane at LHC, assuming $\sqrt{s}=13$ TeV. The invisible branching ratio BR$(\phi_1\rightarrow \psi\psi)$ is shown as dashed lines. The other parameters are fixed as: $g_{\psi}=0.1$, $m_{\psi}=330$ GeV, $c_{2}^{\phi_1}=0$ and $\Lambda_{\phi_1}=3$ TeV.}
\label{fig:resonantprodc3c1}
\end{center}
\end{figure}

In Fig.~\ref{fig:resonantprodc3c1} we depict the contours of the production cross section times the  diphoton branching ratio of $\phi_1$ as well as the invisible branching ratio of $\phi_1$ in the ($c_{1}^{\phi_1}$, $c_{3}^{\phi_1}$) plane, cf.\ Eq.~\eqref{eq:gauge}, assuming $g_{\psi}=0.1$, $m_{\psi}=330$~GeV, $c_{2}^{\phi_1}=0$ and $\Lambda_{\phi_1}=3$~TeV. We can already see that in order to suppress the direct resonant production of $\phi_1$, the product $c_1^{\phi_1}\times c_3^{\phi_1}$ has to be $\lesssim\mathcal{O}(1\times10^{-4})$. Thus, although our numerical scans will cover larger ranges of values for these two parameters, we will discuss our results focusing on this region of the parameter space.  

\subsection{Consistency of the parent resonance framework with LHC constraints}
In what follows, we assume that $\phi_1$ is dominantly produced via Eq.~(\ref{eq:decaychain}). However, as discussed in the Introduction, $\phi_1$ cannot decay dominantly into SM particles since this hypothesis is experimentally disfavoured. As a consequence, we assume that the largest branching ratio of $\phi_1$ is into pairs of DM particles $\psi$, i.e.\ we consider the case where $\Gamma(\phi_1\rightarrow\psi\psi)\sim\Gamma_{\phi_1}$. The additional benefit from this assumption is that a large invisible branching ratio further helps to suppress the signal from direct production of the light pseudoscalar. In the following we fix the branching ratio BR$(\phi_1\rightarrow \psi\psi)=0.9\,(0.8)$ in Scenario 1 (2). With these assumptions, 81$\%\,(64\%)$ of the pair produced $\phi_1$ events decay invisibly, while 18$\%\,(32\%)$ of the events decay into two SM gauge bosons and missing transverse energy and the final state with four SM gauge bosons has a branching ratio of 1$\%\,(4\%)$. Since the differential distributions of the diphoton events have not been published so far, we will probe scenarios with different spectra of missing transverse energy and $p_T$ of the  diphoton system (see also Ref.~\cite{Bernon:2016dow} for a detailed discussion).
If the mass splitting between $\phi_2$ and $\phi_1$ is minimised while allowing for an on-shell decay of $\phi_2\rightarrow\phi_1\phi_1$, the kinetic energy release can be suppressed. For this reason our benchmark points will approximately fulfil this relation between the masses of the two pseudoscalar states:
\begin{equation}
m_{\phi_2}\approx 2\,m_{\phi_1}\,.
\end{equation}
Under this condition, both $\phi_1$ are produced at rest in the $\phi_2$ frame and thus the net transverse missing energy distribution is minimised in the $\gamma\gamma\psi\psi$ final state. However, the photon pair will still have non-vanishing transverse momentum. In addition, due to initial state radiation, the diphoton pair can get an additional boost which could give rise to a harder transverse momentum distribution of the diphoton system.

Since we do not fix the branching ratio BR$(\phi_2\rightarrow\phi_1\phi_1)$, a sizeable branching ratio of $\phi_2$ into SM gauge bosons is possible. As a consequence, dijet~\cite{Aad:2014aqa,Khachatryan:2015sja,Khachatryan:2015dcf,ATLAS:2015nsi} and diphoton signatures from the heavy resonance $\phi_2$ could be observable and we have to check that our scenarios do not violate experimental limits. Another set of constraints comes from monojet and monophoton searches~\cite{Aad:2014tda,Khachatryan:2014rra,Aad:2015zva} and these are explicitly checked in our Monte Carlo simulation as discussed later. Finally, for the decay chain where one of the daughter $\phi_1$ decays to the DM, while the other to jets (perhaps with intermediate gauge bosons) the jets plus missing transverse energy search at $13$~TeV~\cite{ATLAS-CONF-2015-062} is also applied.

\section{Two scenarios for the diphoton excess}

In the effective Lagrangian of Eq.~(\ref{eq:gauge}) there are six dimensionless couplings $c_j^{\phi_i}$ which are {\it a priori} free parameters.
In this Section, we discuss two scenarios - see Table~\ref{tab:scenarios} - where the effective couplings can either be vanishing or they can be related as reminiscence of a more realistic and UV-complete model.

\begin{table}
\centering \renewcommand{\arraystretch}{1.3}
\scalebox{1}{
\begin{tabular}{|c|c|c|}\hline
 & Scenario 1 & Scenario 2 \\
\hline
\hline
$c_1^{\phi_1}$& [$10^{-3}$,\,$1$] & $0.905\,c_3^{\phi_1}$\\ 
\hline
$c_2^{\phi_1}$& $\dfrac{g_2^2}{g_Y^2}\times c_1^{\phi_1}$  & $0.579\,c_3^{\phi_1}$\\ 
\hline
$c_3^{\phi_1}$& 0 & [$10^{-3}$,\,$1$]\\ 
\hline
$c_1^{\phi_2}$& 0 & $0.905\,c_3^{\phi_2}$\\ 
\hline
$c_2^{\phi_2}$& 0 &$0.579\,c_3^{\phi_2}$\\ 
\hline
$c_3^{\phi_2}$& [$10^{-3}$,\,$1$]  &  [$10^{-3}$,\,$1$]\\ 
\hline
$\Lambda_{\phi_1}$& 3 TeV & 3 TeV\\ 
\hline
$\Lambda_{\phi_2}$& 3 TeV & 3 TeV\\ 
\hline
$m_{\phi_1}$& 750 GeV &  750 GeV \\ 
\hline
$m_{\phi_2}$& 1510 GeV &  1600 GeV \\ 
\hline
$\lambda$& [200,\,5000] GeV &[200,\,5000] GeV \\ 
\hline
BR$(\phi_1\rightarrow \psi\psi)$ & 90\% & 80\% \\ \hline
\end{tabular}}
\caption{Definition of the input values and ranges for the parameters of Scenario 1 and Scenario 2. The invisible branching ratio of $\phi_1$ is fixed in both scenarios as specified in the last row.}
\label{tab:scenarios} 
\end{table}

\subsection{Numerical Tools}
The full Lagrangian of Eqs.~(\ref{eq:kin})--(\ref{eq:DM}) was implemented using {\tt FeynRules 2.3.13} \cite{Alloul:2013bka} and an {\tt UFO} output~\cite{Degrande:2011ua} was created for the numerical studies. We generated parton level signal events with {\tt Madgraph 2.3.3}~\cite{Alwall:2014hca} interfaced with {\tt Pythia 6.4}~\cite{Sjostrand:2006za} for the parton shower, underlying event  structure and hadronisation. We have implemented the 8 and 13 TeV diphoton searches from ATLAS and
CMS~\cite{ATLASdiphoton2015,CMSdiphoton2015,Aad:2014ioa,Aad:2015mna,Khachatryan:2015qba,CMS-PAS-EXO-12-045}
into the {\tt CheckMATE 1.2.2} framework \cite{Drees:2013wra} with its
{\tt AnalysisManager} \cite{Kim:2015wza}.  
{\tt CheckMATE 1.2.2} is based on the fast detector simulation {\tt
  Delphes 3.10} \cite{deFavereau:2013fsa} with heavily  
modified detector tunes and it determines the number of expected signal events passing the
selection cuts of the particular analysis. The selection cuts  
for both ATLAS and CMS 13~TeV diphoton analyses are shown in Table~\ref{tab:selection13tev}. The resulting signal efficiency varies between 20$\%$ and 60$\%$ depending on the signal region, the experiment and the centre-of-mass energy. The analyses were validated to reproduce efficiencies reported by ATLAS and CMS. Finally, experimental constraints from dijet searches, jets and missing transverse momentum~\cite{ATLAS-CONF-2015-062} and monojet~\cite{Aad:2014tda,Khachatryan:2014rra,Aad:2015zva} searches have been implemented into {\tt CheckMATE 1.2.2} and have been fully validated against public results.
  
\begin{table}
\centering \renewcommand{\arraystretch}{1.3}
\scalebox{1}{
\begin{tabular}{|c|c|}\hline
 ATLAS & CMS \\
\hline
\hline
$p_T(\gamma)\ge$25 GeV  &
$p_T(\gamma)\ge$75 GeV \\ 
\hline
$|\eta^{\gamma}|\le2.37$ & $|\eta^{\gamma}|\le 1.44$
 or $1.57 \le |\eta^{\gamma}| \le 2.5$ 
\\ 
& at least one $\gamma$ with $|\eta^{\gamma}|\le1.44$\\
\hline
$E_T^{\gamma_1}/m_{\gamma\gamma}\ge0.4$,
$E_T^{\gamma_2}/m_{\gamma\gamma}\ge0.3$ & $m_{\gamma\gamma}\ge230$ GeV \\ \hline
\end{tabular}}
\caption{Selection cuts of the 13 TeV ATLAS/CMS diphoton searches
  \cite{ATLASdiphoton2015,CMSdiphoton2015}. 
\label{tab:selection13tev} }
\end{table} 
 
\subsection{Scenario 1}
\subsubsection{Benchmark Parameters}
In this scenario we achieve dominant indirect production simply by setting the effective coupling between the gluons and $\phi_1$  to zero, $c_3^{\phi_1}=0$\footnote{This choice of the effective couplings seems arbitrary, but it resembles the structure of a hierarchical scenario in a simple composite Higgs model discussed in Ref.~\cite{Harigaya:2015ezk}.}. As a consequence, the lighter pseudoscalar cannot be produced in gluon fusion and the production mechanism via photon-photon collision is heavily suppressed. Hence, $\phi_1$ has to be produced in the cascade decay of the heavy parent resonance. The solely allowed coupling between the heavy resonance $\phi_2$ and the SM gauge bosons is to gluons.  
We fix $m_{\phi_2}=1510$ GeV thus minimising the missing transverse momentum of the photon pair. Motivated by the DM constraints, see Section~\ref{sec:dm}, in the following we closely analyse a parameter point with the couplings of $\phi_1$ defined as follows:
\begin{equation}
 c_1^{\phi_1} = 9.3\cdot 10^{-3}\,,\, g_\psi = 1.24\cdot 10^{-1}\,,\, m_\psi = 337~\mathrm{GeV}\,.
\end{equation}
The other parameters are summarised in Table~\ref{tab:scenarios}. 
The invisible branching ratio of the light pseudoscalar is  90\% while the one into photons is 0.7\%. The lighter pseudoscalar couples to the EW SM gauge bosons, namely $WW$, $ZZ$, $Z\gamma$ and $\gamma\gamma$ states with the following ratios of the partial decay widths of $\phi_1$:
\begin{equation}
\gamma Z/\gamma\gamma=0.73,\quad WW/\gamma\gamma=8.4,\quad ZZ/\gamma\gamma=3.9\,.
\end{equation}
\subsubsection{Constraints}
Since the light pseudoscalar does not couple to gluons, its production cross section is very small and does not affect the phenomenology at the LHC. In particular, we do not have to worry about diphoton constraints due to the gluon-initiated production. The smallness of the BR($\phi_1 \to \gamma \gamma$) further assures that the diphoton signal from photon-photon fusion production of  $\phi_1$ is also negligible. On the other hand, the couplings of the light pseudoscalar are constrained by the astrophysical observables as we will see in Section~\ref{sec:dm}.  As to the heavy pseudoscalar, whose coupling to gluons is non-vanishing, one has to consider constraints coming from dijet spectra at $m_{jj} \sim 1500$~GeV in the 8 and 13~TeV data.

\begin{figure}[!t]
 \begin{center}
  \includegraphics[width=0.49\textwidth]{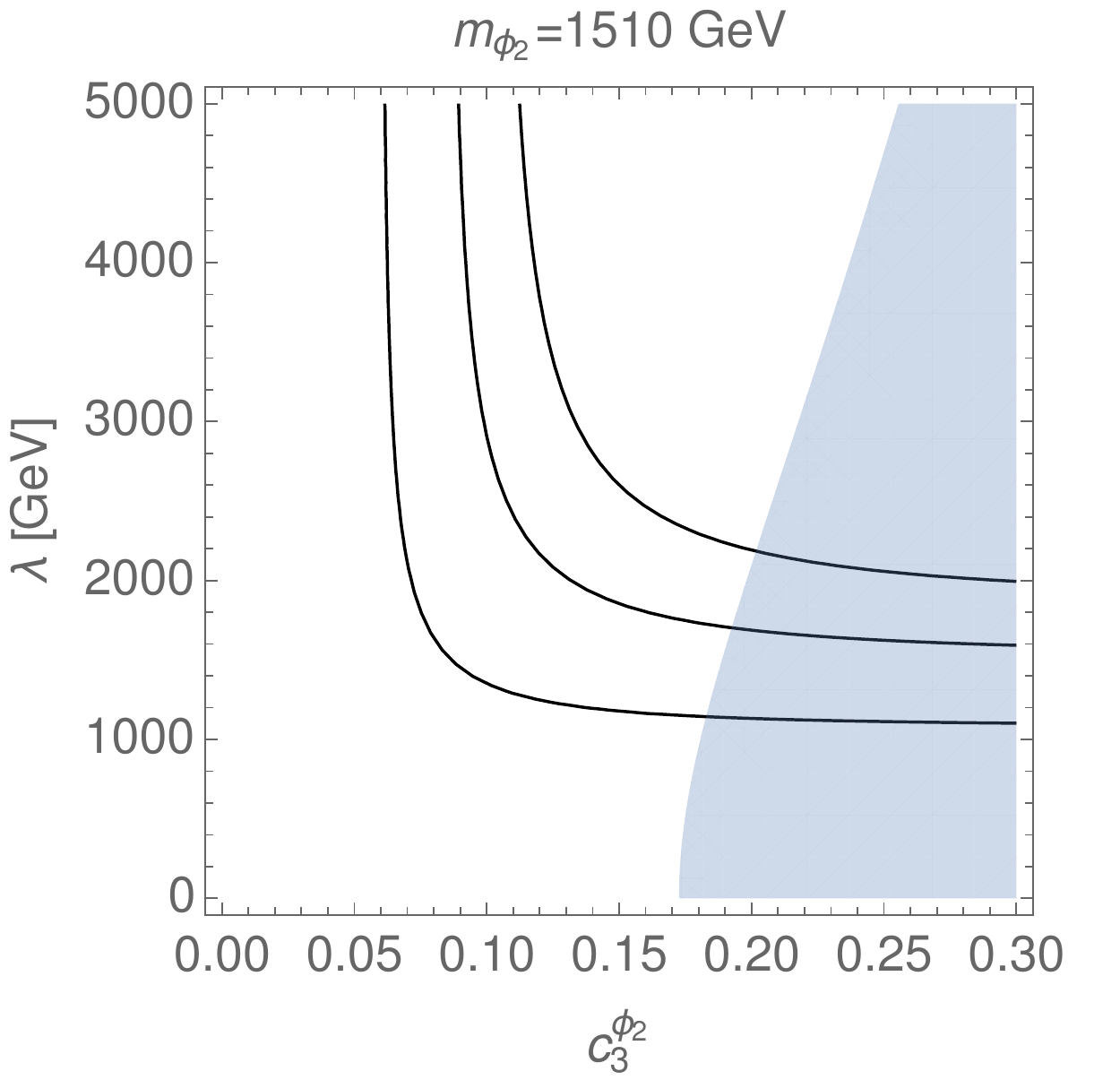} 
 \end{center}
\caption{Cross section contours (from bottom: 3, 6, 9~fb) for $\sigma(pp\to \psi\psi\gamma\gamma)$ for $m_{\phi_2}=1510$~GeV in  Scenario 1 assuming BR$(\phi_1 \to \psi\psi) = 90\%$ and BR$(\phi_1 \to \gamma\gamma) = 0.7\%$. The shaded area is excluded by dijet production.}
\label{fig:1510c3lambda}
\end{figure} 

\subsubsection{Results}
In Fig.~\ref{fig:1510c3lambda} we show the cross section contours for $\sigma(pp\rightarrow\psi\psi\gamma\gamma)$ for Scenario 1 with $m_{\phi_2}=1510$~GeV. The light blue shaded area is excluded by dijet searches~\cite{Aad:2014aqa,Khachatryan:2015sja,Khachatryan:2015dcf,ATLAS:2015nsi}. The contours correspond to the cross sections of 3, 6, 9~fb for the production of the diphoton final state, which translates to $\sim$5, 10 and 15 events in the mass window 700--800~GeV. The simulated efficiency is $\sim 75\%$ for ATLAS~\cite{ATLASdiphoton2015}. The diphotons have relatively low momentum and are very central in the detector which results in a very low contribution - consistent with the data - to the CMS EBEE~\cite{CMSdiphoton2015} (ECAL barrel--end-cap) signal region, 1.0, 2.1 and 3.3 events respectively. The yield in the barrel signal region is similar to that of ATLAS.

\begin{figure}
 \begin{center}
  \includegraphics[width=0.5\textwidth]{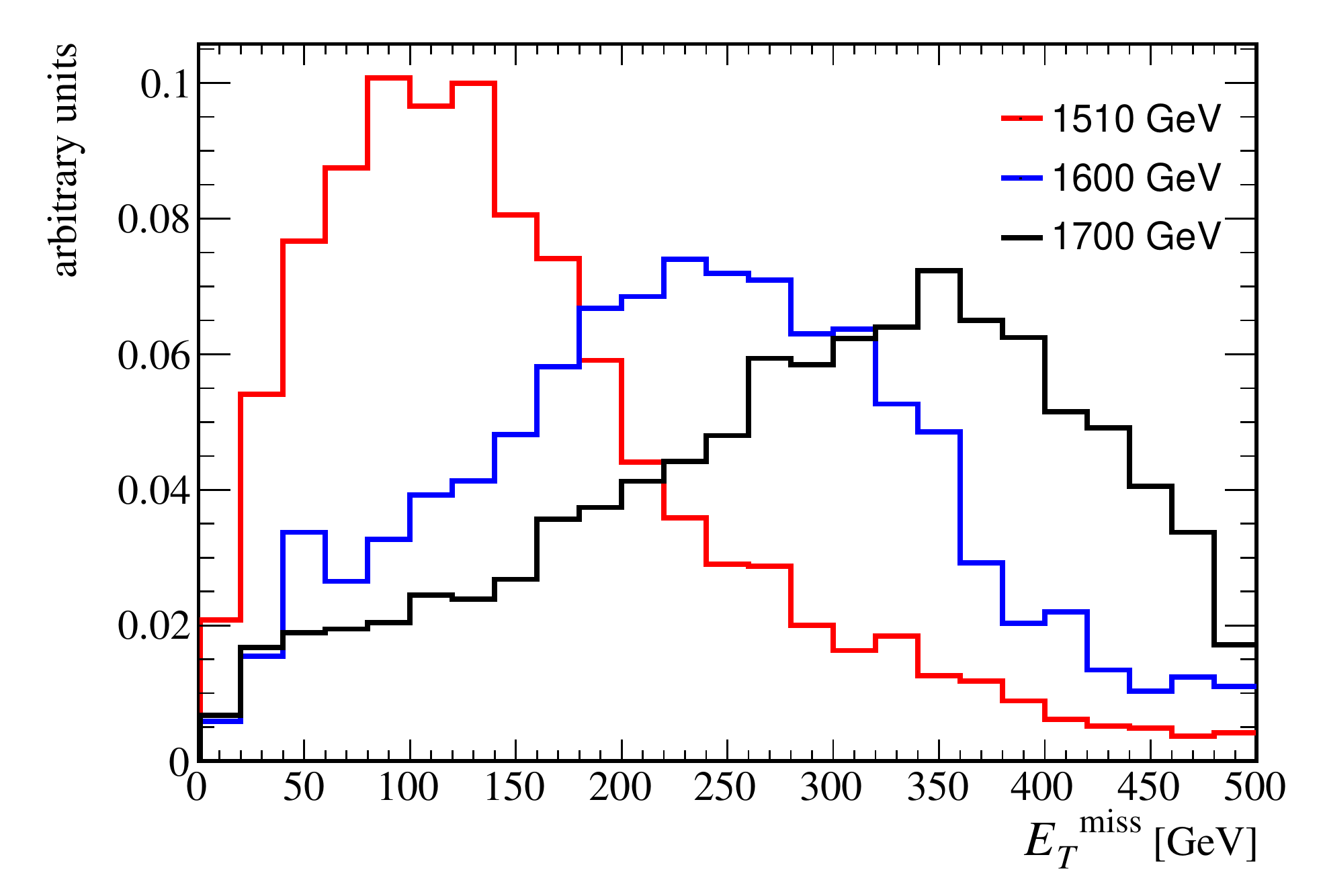} 
 \end{center}
\caption{Missing transverse energy distribution of the process $\sigma(pp\to \psi\psi\gamma\gamma)$ for $m_{\phi_2}=1510,\,1600,\, 1700$~GeV in Scenario~1 -- red, blue and black histograms, respectively.}
\label{fig:met}
\end{figure}

In Fig.~\ref{fig:met} we show the missing transverse energy distribution for different heavy scalar masses, $m_{\phi_2} = 1510$, 1600 and 1700~GeV.
We can see that even for the mass degenerate scenario, the net transverse missing energy is not negligible and the distribution peaks around 100--150~GeV. As expected, once the mass gap between $\phi_2$ and $\phi_1$ increases, the distribution shifts to the right. This distinctive feature can be used to measure the mass of the heavy scalar once the signal is confirmed and with higher statistics (see also the detailed discussion in Ref.~\cite{Bernon:2016dow}).
 
\subsection{Scenario 2}
\subsubsection{Benchmark parameters}
In this scenario, we assume that all the effective couplings $c_j^{\phi_i}$ in Eq.~(\ref{eq:gauge}) are non-vanishing. For simplicity, we fix the relations between the various $c_j^{\phi_i}$
as in scenario~{\it F1} of Ref.~\cite{Knapen:2015dap}. The authors introduced heavy vector-like fermions with the following SM gauge group SU(3)$_C\times$SU(2)$_L\times$U(1)$_Y$ assignment: $(3,2,7/6)$. The tree level decays of the pseudoscalar into these new vector-like fermions are kinematically closed and thus the pseudoscalar only decays into the SM gauge bosons via loop induced couplings as well as into DM. 
We assume that the following parameters define the physics of the light pseudoscalar:
\begin{equation} \label{eq:sc2par}
c_3^{\phi_1} = 1.4\cdot 10^{-2}\,,\, g_\psi = 6.6\cdot 10^{-2}\,,\, m_\psi = 341~\mathrm{GeV}\,.
\end{equation}
This choice of parameters gives the correct DM relic density as discussed in the following Section. The branching ratio of the light pseudoscalar into DM is 80\% while  BR($\phi_1 \to \gamma \gamma) \sim$ 1.4\%.  The remaining couplings, $c_1^{\phi_1}$ and $c_2^{\phi_1}$, are related to $c_3^{\phi_1}$, as shown in Table~\ref{tab:scenarios}. 
Moreover, we have increased the mass of $\phi_2$ to $m_{\phi_2}=1600$ GeV (for the reason explained in the next paragraph).
The ratios of the partial widths of $\phi_1$ into SM gauge bosons are given by:
\begin{equation}
\gamma Z/\gamma\gamma=0.06,\quad WW/\gamma\gamma=0.91,\quad ZZ/\gamma\gamma=0.6,\quad gg/\gamma\gamma=11.62\,.
\end{equation}

\subsubsection{Constraints}
In this scenario where both scalars couple to gluons, the constraints from dijet searches~\cite{Aad:2014aqa,Khachatryan:2015sja,Khachatryan:2015dcf,ATLAS:2015nsi} and  diphoton searches have to be taken into account for both invariant masses of 750 and 1600~GeV. However, the dominant branching ratio for both $\phi_1$ and $\phi_2$ is not the one into jets. We have checked that for the above choice of parameters, Eq.~\eqref{eq:sc2par}, the dijet constraints for $\phi_1$ are easily fulfilled. The situation is more tricky for the diphoton final state. In fact, since $\phi_1$ couples to gluons, direct production of the lighter resonance $\phi_1$ is now possible and thus we have to check that the resonant production of $\phi_1$ is still suppressed. We found that $\sigma(pp\to\phi_1\to\gamma\gamma) \simeq 2$~fb at $\sqrt{s} = 13$~TeV and one can indeed expect $\sim 3$ events in the signal region at 8 and 13~TeV. 

Similarly for $\phi_2$, the dijet constraints have been checked and we have found that they become relevant for values of the coupling $c_3^{\phi_2} \gtrsim 0.2$. There is also a not-so-welcome contribution to the diphoton final state at $m_{\gamma\gamma}=1600$~GeV. This one will turn out to be a much stronger constraint. Incidentally, there are two events in the ATLAS search~\cite{ATLASdiphoton2015} at this invariant mass. The expected number of background events is $\sim 0.8$ for $m_{\gamma\gamma} > 1500$~GeV. Since both events are located closely together one can speculate that they originate from a hypothetical new particle and under this assumption the local significance is $\sim 3\sigma$. This motivates our choice of $m_{\phi_2}=1600$~GeV.\footnote{Note that choosing any other mass would actually lead to even stronger combined exclusion limits from both experiments at the high invariant masses.} In any case, we also have to take into account the bound from CMS EBEB (ECAL barrel) signal region, where no events were observed at $m_{\gamma\gamma} = 1600$~GeV. Since the diphotons have the largest branching ratio from the EW gauge bosons, one clearly sees that the constraints from other diboson production processes are easily fulfilled~\cite{Chatrchyan:2012rva,Aad:2015kna,Aad:2015ipg}. Nevertheless one could eventually expect to observe e.g.\ $ZZ$ resonant production in the 4-lepton channel.  

\subsubsection{Results}
\begin{figure}
 \begin{center}
  \includegraphics[width=0.49\textwidth]{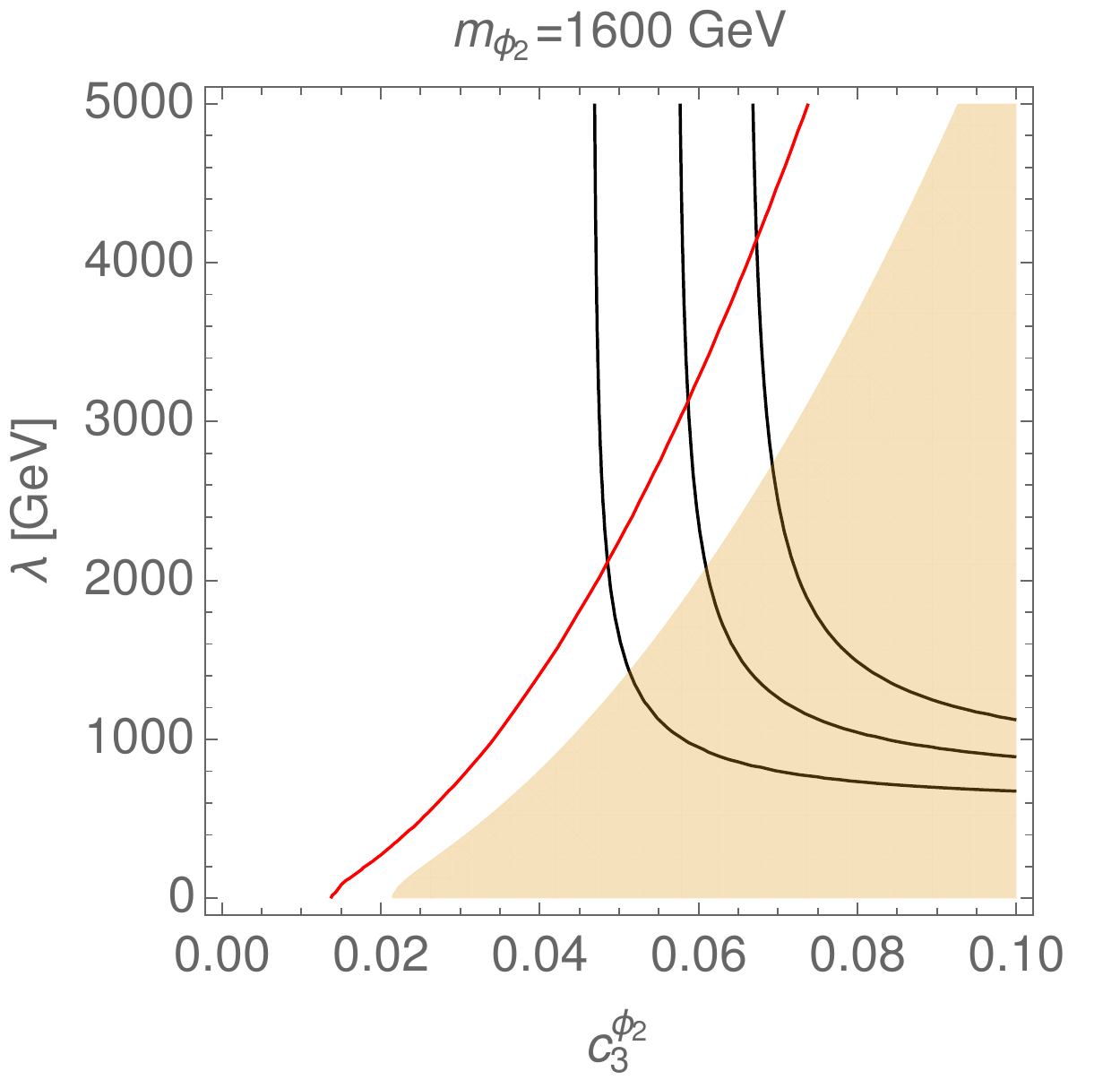} 
 \end{center}
\caption{Cross section contours (from bottom: 2.6, 3.9, 5.2~fb) for $\sigma(pp\to \psi\psi\gamma\gamma)$ for $m_{\phi_2}=1600$~GeV in Scenario 2, assuming BR$(\phi_1 \to \psi\psi) = 80\%$ with BR$(\phi_1 \to \gamma\gamma) = 1.4\%$. The light shaded area is excluded by the direct diphoton production $\phi_2 \to \gamma \gamma$ at $m_{\gamma\gamma} =1600$~GeV. The red line corresponds to the best fit from two high mass events (at 1600 GeV) in ATLAS. }
\label{fig:1600c3lambda}
\end{figure}

In Fig.~\ref{fig:1600c3lambda} we present the cross section contours for $\sigma(pp\rightarrow\psi\psi\gamma\gamma)$ in Scenario 2. In the shaded area, the resonant diphoton production via $\phi_2$ violates experimental bounds, mainly from the CMS EBEB signal region. The red line corresponds to the expected observation of $1.2$ events at $m_{\gamma\gamma} = 1600$~GeV. 
The cross section contours 2.6, 3.9, 5.2~fb correspond to the expected 6, 9 and 12 events, respectively. As discussed in the previous paragraph, the additional contribution of 3 events due to the resonant $\phi_1$ production would be expected. Similarly to Scenario~1, the diphotons have relatively low momentum and are very central in the detector, which results in a very low contribution, consistent with the data, to the CMS EBEE~\cite{CMSdiphoton2015} signal region, 1.5, 2.3 and 3.1 events respectively. The combination of the photon constraints at the high mass and at 750~GeV, narrows the preferred parameter space to $c_3^{\phi_2} \sim \mathcal{O}(0.05)$ and $\lambda \gtrsim 2.5$~TeV. We note that the ratio  of the anomalous couplings of both scalars is approximately given by $c_3^{\phi_2}/c_3^{\phi_1} \sim 3$--$4$.

\begin{figure}
 \begin{center}
  \includegraphics[width=0.49\textwidth]{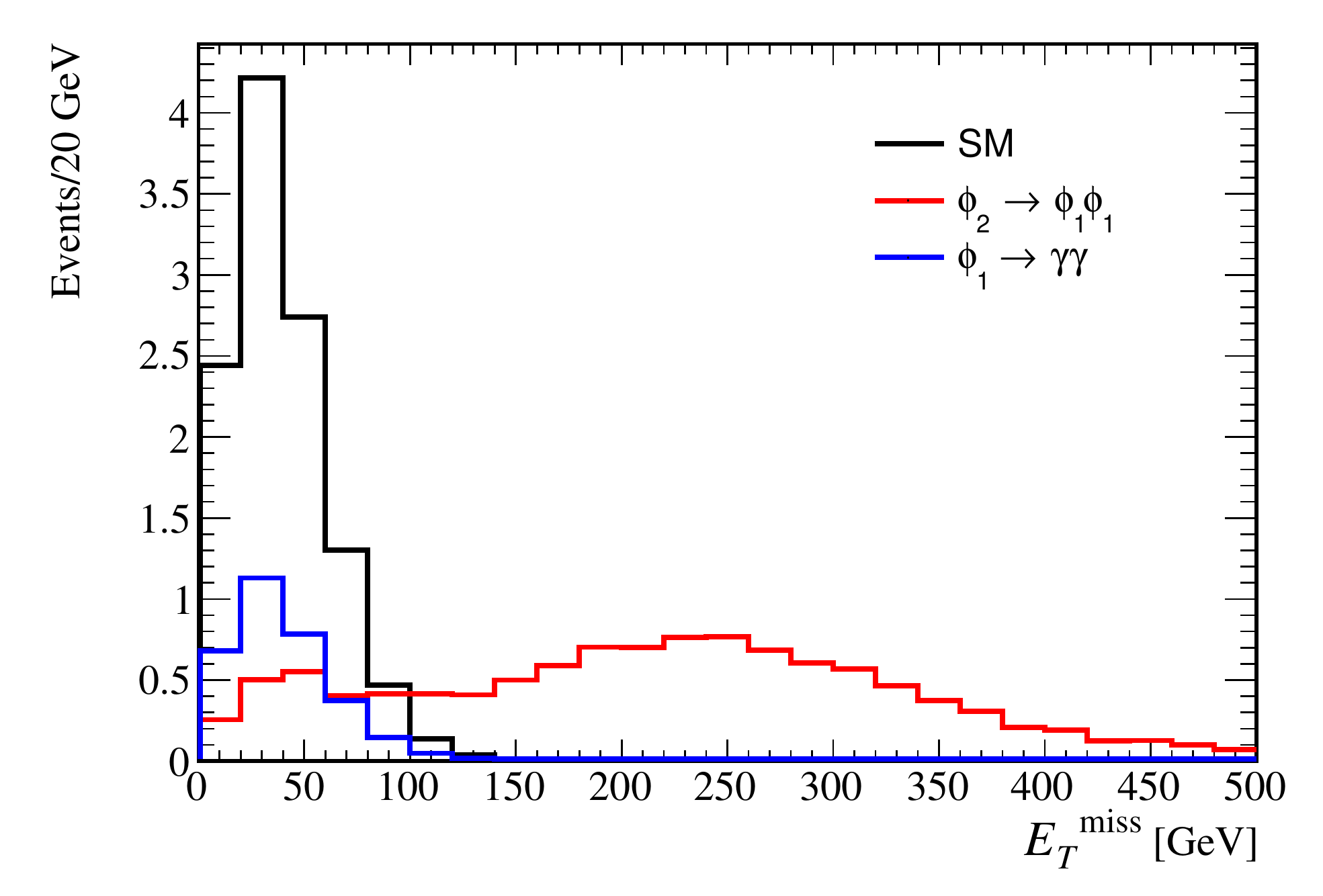} 
  \includegraphics[width=0.49\textwidth]{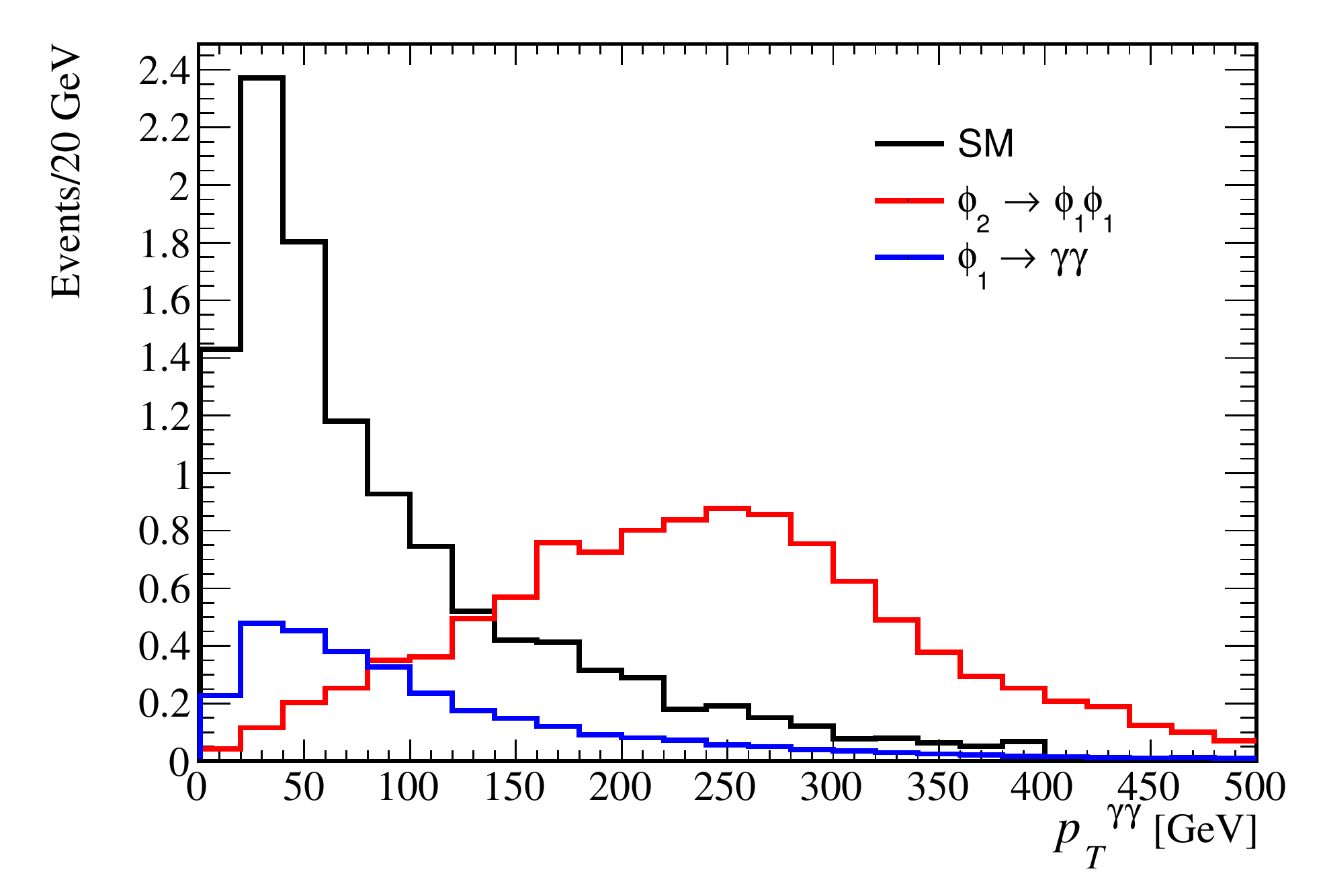} 
 \end{center}
\caption{Missing transverse energy distribution (left) and transverse momentum of the photon pair (right) of the process $\sigma(pp\to \gamma\gamma + X)$ for $m_{\phi_2}=1600$~GeV in Scenario~2. The black histogram is for the SM background, the red for the full decay chain, Eq.~\eqref{eq:decaychain}, and the blue for the direct production of the light scalar, Eq.~\eqref{eq:phi1}.}
\label{fig:metF1}
\end{figure}

Figure~\ref{fig:metF1} shows the missing transverse energy distribution and transverse momentum distribution of the photon pair. We compare the expectation for the SM background simulated with \texttt{MadGraph} and normalised to the observed number of events. Both new physics contributions from $pp \to \phi_1 \to \gamma\gamma$ and $pp\to \phi_2 \to \phi_1 \phi_1$ are shown separately.  While the light $\phi_1$ production exhibits a shape similar to the background, the contribution due to $\phi_2$ is heavily shifted towards higher values, as already observed in Scenario~1. This provides a unique feature of the model studied in this paper.

\section{Dark Matter phenomenology}\label{sec:dm}
In this Section we discuss the possibility that the lightest pseudoscalar mediates the interactions of a DM candidate -- a Majorana fermion -- with the SM.
While several aspects of the phenomenology of DM with a possible 750 GeV (pseudo)scalar mediator have already been studied in the literature \cite{DM750}, we investigate the DM in a novel setup and taking into account LHC constraints.

Within this framework, the DM particles $\psi$ annihilate into SM particles via the $s$-channel exchange of the pseudoscalar mediator $\phi_1$. The final state particles which can be produced in the annihilations are: $\psi \psi \rightarrow \gamma \gamma$, $g g$, $\gamma Z$, $Z Z$ and $W^+ W^-$ depending on the couplings and on $m_\psi$. We compute the relic density of $\psi$ using the code {\tt micrOMEGAs} 4.1 \cite{Belanger:2014vza}, for which we have implemented our model in the {\tt CalcHEP} format \cite{Belyaev:2012qa}. 

To address the phenomenology of DM in the parameter space allowed by our model, we have performed a random scan with 10000 points over the parameters $(m_\psi, c^{\phi_i}_1)$. The other parameters are fixed by imposing the relations in Tab.~\ref{tab:scenarios} and the invisible branching ratio BR($\phi_1 \rightarrow \psi \psi$) = 90 (80) \% in Scenario 1 (2). Fixing the BR($\phi_1 \rightarrow \psi \psi$) leads to a relation between $g_\psi$ and $c^{\phi_1}_1$. We have limited the DM mass up to $m_{\phi_1}/2$, in order to allow for invisible decays of $\phi_1$ as already discussed in Section \ref{sec:model}.

\subsubsection*{Relic density} 
We compute the relic abundance of the DM as a function of its mass $m_\psi$ and coupling $g_\psi$. We apply the current constraints from the PLANCK satellite \cite{Ade:2013zuv} with the best fit value of the relic density corresponding to $\Omega_{\psi} h^2 =(0.1198 \pm 0.0026)$ \cite{Agashe:2014kda}. We consider a benchmark point consistent with the PLANCK bound, if the computed relic density does not exceed the measured abundance. As a consequence, we regard under-abundant DM as cosmologically safe although additional DM candidates have to be introduced in the context of the standard cosmology.

\subsubsection*{Indirect detection} 
The observation of the final products of DM annihilation is a promising method to search for DM. In the scenarios considered here, $\phi_1$ does not couple to fermions, therefore the possible contributions to the velocity averaged annihilation cross section are: $\langle \sigma v \rangle_{tot} = \langle \sigma v \rangle_{\gamma \gamma} +\langle \sigma v \rangle_{WW}+\langle \sigma v \rangle_{Z \gamma}+\langle \sigma v \rangle_{ZZ}$ in Scenario 1 and $\langle \sigma v \rangle_{tot} = \langle \sigma v \rangle_{\gamma \gamma} +\langle \sigma v \rangle_{Z \gamma}+\langle \sigma v \rangle_{ZZ}+ \langle \sigma v \rangle_{gg}$ in Scenario 2.

Among all possible SM particles which can be produced by DM annihilation, photons are among the most powerful messengers for the ID of DM, since they proceed almost unperturbed when propagating through the Universe. $\gamma$-rays from DM annihilation can be produced via a variety of mechanisms. Here we have two different $\gamma$-ray signatures. Firstly, the DM annihilation into the SM gauge bosons, $\psi \psi \rightarrow ZZ$, $Z \gamma$, $W^+ W^-$ and $gg$ which eventually hadronise and/or decay producing lighter mesons ($\pi$) that give rise to a continuous spectrum. Secondly,
both our scenarios are characterised (owing to the connection to the diphoton signal at the LHC) by the presence of a monochromatic  $\gamma$-ray signal at $m_\psi$.
Since no DM signal has been found so far by ID experiments, we apply the latest bounds from the Fermi-LAT collaboration on the DM annihilation cross sections \cite{Ackermann:2015zua,Ackermann:2015lka}. 

In order to compare with the experimental bounds from ID, we rescale the DM annihilation cross section taking into account the ratio of the value of the relic density computed in our scenarios and the observed one. 
We impose the limits on the continuous spectrum from the latest observation of dwarf spheroidal galaxies (dSphs) of the Milky Way made by Fermi-LAT \cite{Ackermann:2015zua}.  
For Scenario 1, we compare the experimental bounds from he $W^+ W^-$ final state provided by Fermi-LAT \cite{Ackermann:2015zua} with our predicted annihilation cross section $\langle \sigma v \rangle_{ZZ} + \langle \sigma v \rangle_{W^+ W^-} + \frac{\langle \sigma v \rangle_{Z \gamma}}{2}$. In Scenario 2 we compare the experimental bounds for the $u \bar{u}$ channel obtained by the Fermi-LAT collaboration \cite{Ackermann:2015zua} with our predicted annihilation cross section $\langle \sigma v \rangle_{gg}$\footnote{For this purpose, we notice that the $\gamma$-ray spectra from light quarks and gluons is similar as well as the $\gamma$-ray spectra from gauge bosons is almost universal, as advocated for instance in Ref.~\cite{Cirelli:2010xx}.}.

We further consider the limits on the annihilation cross section from Galactic $\gamma$-ray line searches from Fermi-LAT. In this case, we compare the  experimental bounds from  Ref.~\cite{Ackermann:2015lka} with the predicted  annihilation cross section $\langle \sigma v \rangle_{\gamma \gamma} + \frac{\langle \sigma v \rangle_{Z \gamma}}{2}$. We consider the limits given by the Fermi-LAT collaboration, both assuming a Navarro-Frenk-White (NFW) profile and an Einasto profile of the spatial distribution of DM in our Galaxy.

\subsubsection*{Direct detection} 
The limits from direct searches for DM are not relevant in the case of a pseudoscalar mediator. The DM-nucleons scattering cross-section is indeed strongly suppressed by the square of the nuclear recoil energy, which is small because of the non-relativistic nature of the interaction (see for instance the discussion in \cite{Boehm:2014hva}). Therefore we do not discuss bounds from direct detection experiments.

\subsubsection*{Numerical results} 
\begin{figure}[!hbtp]
\centering
\includegraphics[width=.55\textwidth]{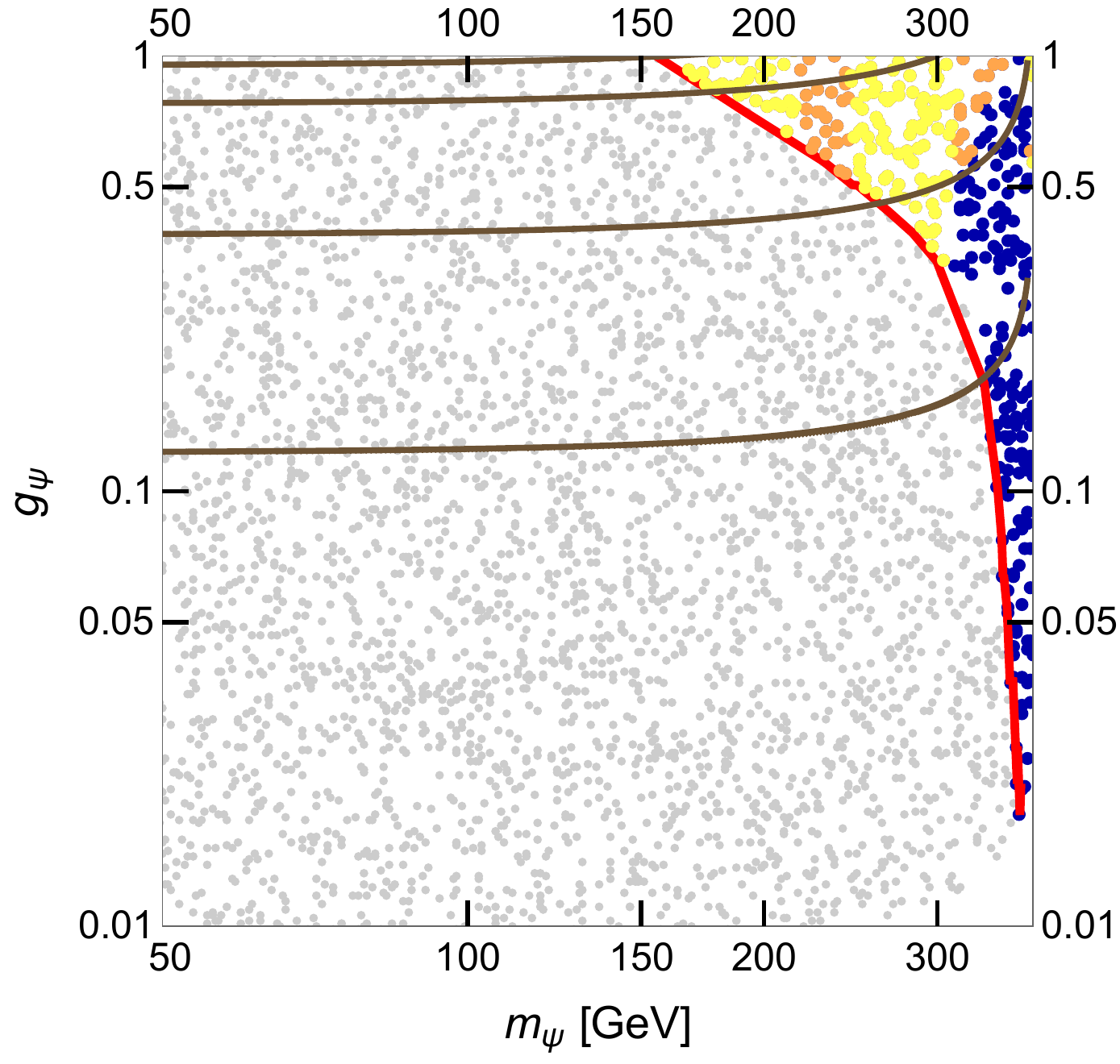}
\caption{\label{fig:gpsi_mpsi_scen1} Results of the numerical scan in the parameter space ($m_\psi, g_\psi$) for Scenario~1 (see Table \ref{tab:scenarios}). Grey points correspond to over-abundant DM and they are excluded by the relic density measurement made by the PLANCK satellite \cite{Ade:2013zuv}. The solutions lying on the red curve have the correct relic density. Blue points correspond to under-abundant DM and are allowed by ID. Yellow (yellow + orange) points are excluded by $\gamma$-ray line searches \cite{Ackermann:2015lka} assuming a NFW (Einasto) profile of the spatial distribution of the DM in our Galaxy. The brown curves indicate different values of the total decay width of $\phi_1$, $\Gamma_{\phi_1} = 1$, 10, 40 and 60~GeV, respectively, from bottom to top. }
\label{fig:DMscenI}
\end{figure}

The results of the numerical scan for Scenario 1 are shown in Fig.~\ref{fig:DMscenI} in the ($m_\psi, g_\psi$) plane. In this scenario, we have fixed  $c_3^{\phi_1} = 0$, BR($\phi_1 \rightarrow \psi \psi) \sim 90$ \% and $\Lambda_{\phi_1}= 3$ TeV,  as described in Section \ref{sec:diphoton} (see Table \ref{tab:scenarios}). We depict as grey points the solutions with $\Omega_{\psi} h^2 \gsim 0.1198$, hence excluded by the relic density measurement made by the PLANCK satellite \cite{Ade:2013zuv}. The solutions lying on the red curve have the relic density $\Omega_{\psi} h^2 \sim (0.1198 \pm 2\sigma)$ while the blue points correspond to under-abundant DM ($\Omega_{\psi} h^2 \lesssim 0.1198$) and are in agreement with ID bounds. Yellow (yellow + orange) points are excluded by $\gamma$-ray line searches \cite{Ackermann:2015lka} assuming a NFW (Einasto) profile of the spatial distribution of the DM in our Galaxy. The $\gamma$-ray line constraints also apply to the points along the red line close to the yellow region. Finally, the brown curves denote the contours of the total $\phi_1$ decay width corresponding to $\Gamma_{\phi_1} = 1$, 10, 40 and 60~GeV (from bottom to top), respectively.

\begin{figure}
\hspace*{-11mm}
\begin{tabular}{cc}
\includegraphics[width=.55\textwidth]{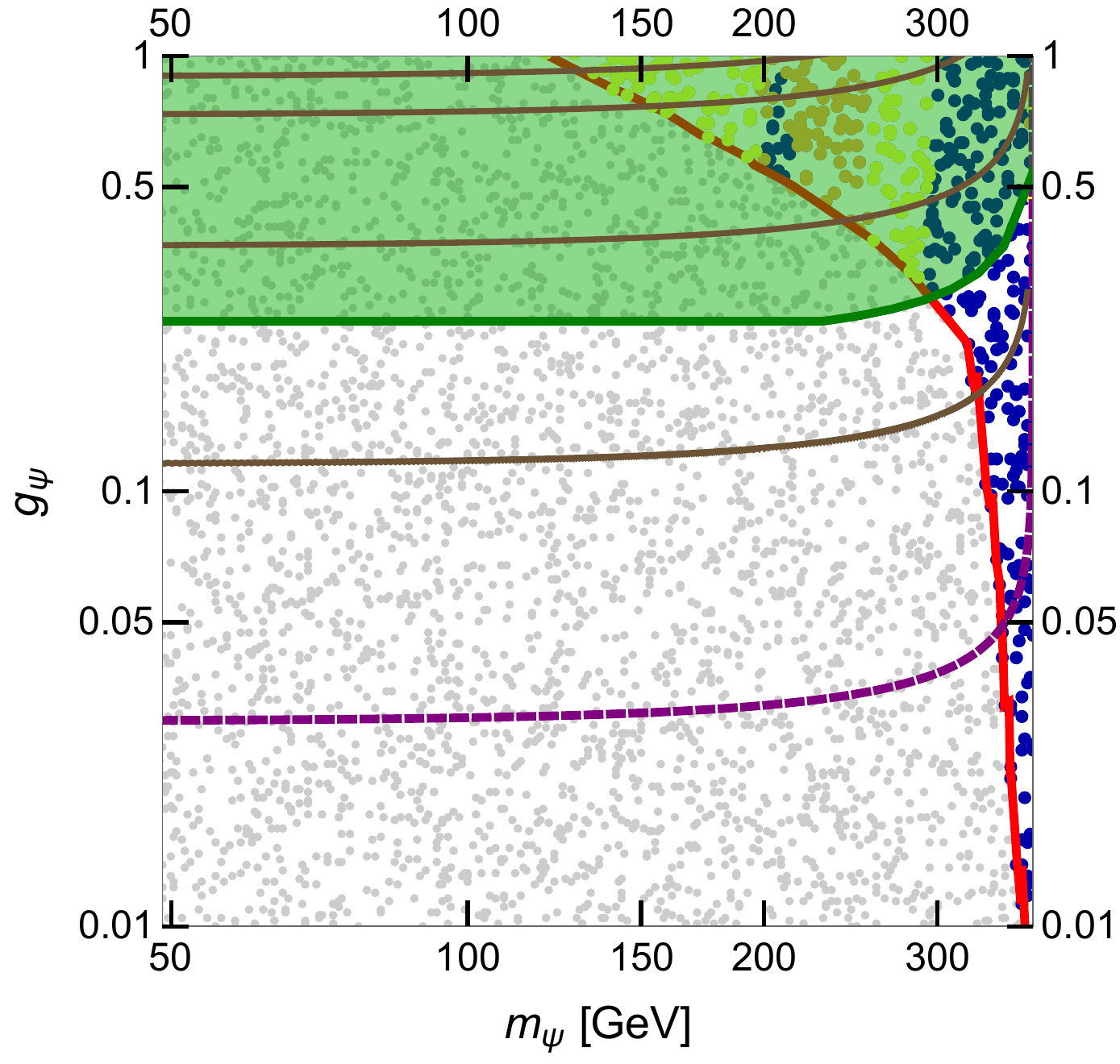}
\hspace*{-3mm}
&\hspace*{-3mm}
\includegraphics[width=.57\textwidth]{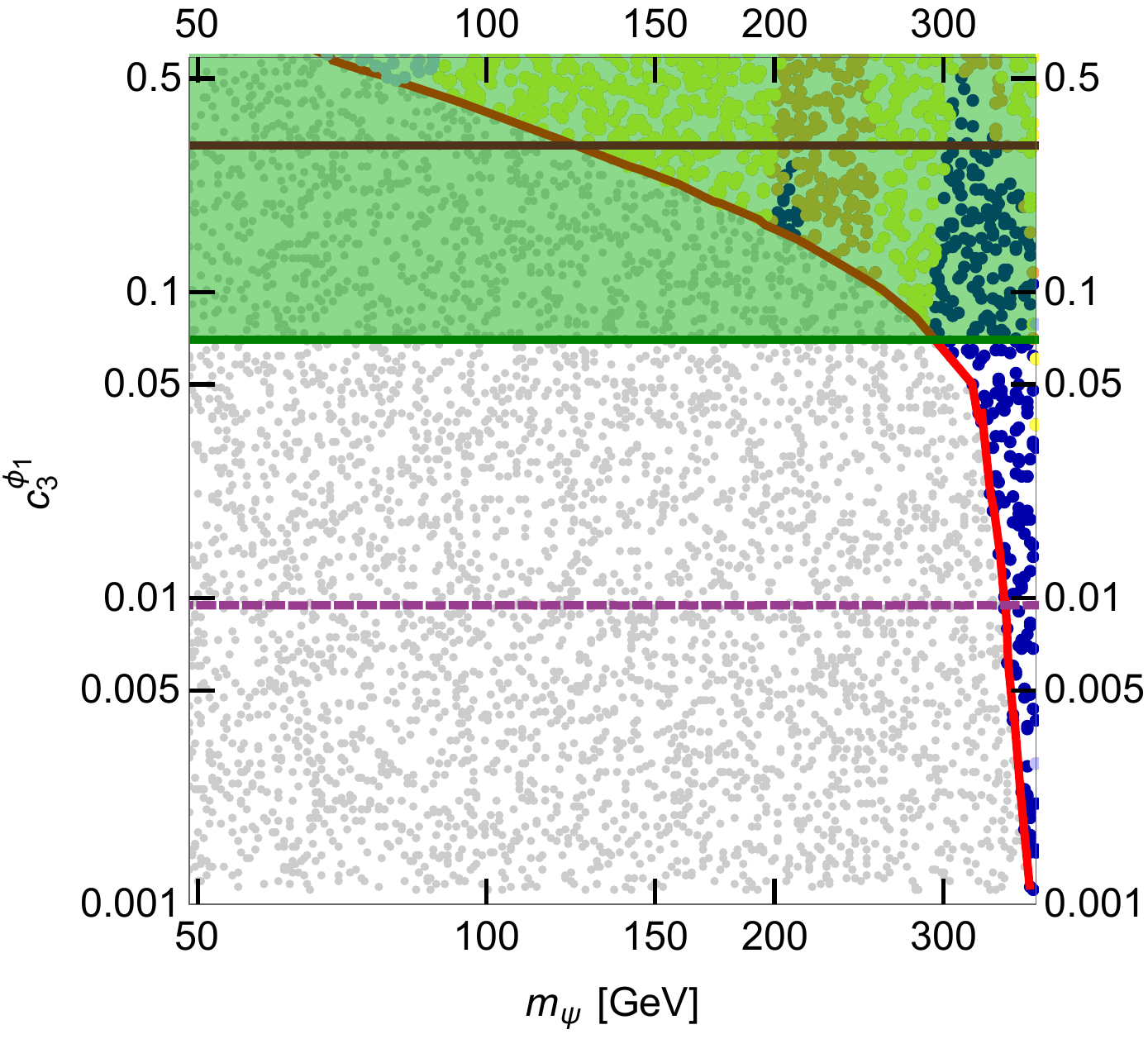} 
\end{tabular}
\caption{Results of the numerical scan in the parameter space ($m_\psi, g_\psi$) (left) and ($m_\psi, c_3^{\phi_1}$) (right) for Scenario~2 (cf.\ model {\it F1} in \cite{Knapen:2015dap} and Table \ref{tab:scenarios}). Grey points are excluded by the relic density measurement by the PLANCK satellite \cite{Ade:2013zuv}. The solutions lying on the red curve give the correct relic density. Blue points correspond to under-abundant DM and are allowed by ID. Light blue points are in disagreement with the latest observation of dSphs \cite{Ackermann:2015zua}.  Yellow (yellow + orange) points are excluded by $\gamma$-ray line searches \cite{Ackermann:2015lka}, assuming a NFW (Einasto) profile of the spatial distribution of the DM in our Galaxy. The green thick line (and the green shaded area above) indicates the upper bound at 95\% C.L.\ on the $pp$ cross section for final states with one energetic jet and large missing transverse momentum at $\sqrt{s}= 8$ TeV from the ATLAS collaboration \cite{Aad:2015zva}. In the panel on the left, the brown curves indicate different values of the total decay width of $\phi_1$, $\Gamma_{\phi_1} = 1$, 10, 40 and 60~GeV, respectively, from bottom to top. In the right panel, the thick brown line at $c_3^{\phi_1} \sim 0.3$ denotes $\Gamma_{\phi_1} = 60$ GeV. The purple dashed line denotes $\sigma(pp\rightarrow \phi_1\rightarrow\gamma\gamma)=1\,$fb. }
\label{fig:DMscenII}
\end{figure}

Concerning Scenario 2, the results of the numerical scan are shown in the two panels of Fig.~\ref{fig:DMscenII}.  The scan ranges of the input parameters are given in Table~\ref{tab:scenarios}. In the left panel, we show the parameter space in the ($m_\psi, g_\psi$) plane. The colour code is the same as in Fig.~\ref{fig:DMscenI}. In the right panel, we depict the results in the ($m_\psi, c_3^{\phi_1}$) plane. 
On top of the relic density and the $\gamma$-ray lines constraints, in this plot we additionally present the collider constraints. 

The green thick line (and the green shaded area above) indicates the upper bound at 95\% confidence level (C.L.) on the $pp$ cross section for final states with one energetic jet and large missing transverse momentum at $\sqrt{s} = 8$~TeV from the ATLAS collaboration \cite{Aad:2015zva}. This bound is placed at $c_3^{\phi_1} \sim 0.07$ and $g_\psi \sim 0.25$--$0.5$ in the two plots of Fig.~\ref{fig:DMscenII}.
Although constraints from the dijet search at $\sqrt{s}= 8$~TeV are not shown in the plots, we have also computed the upper limit at 95\% C.L.\ taken from Ref.~\cite{Franceschini:2015kwy}, which corresponds to $c_3^{\phi_1} \sim 0.14$ and $g_\psi \sim 1$. 

In the left panel of Fig.~\ref{fig:DMscenII}, the thick brown lines denote $\Gamma_{\phi_1} = 1$, $10$, $45$ and 60~GeV (from bottom to top), respectively, while in the right panel, the single brown line corresponds to $\Gamma_{\phi_1} = 60$ GeV at $c_3^{\phi_1} \sim 0.3$. 
 Finally, one has to take into account the direct resonant production of $\phi_1$ followed by the decay to a photon pair. This has to be combined with the non-direct production to obtain a meaningful limit, as detailed in Section~\ref{sec:diphoton}. For guidance we show the purple dashed line corresponding to $\sigma(pp\rightarrow \phi_1\rightarrow\gamma\gamma)=1\,$fb.

In both scenarios, the constraint on the relic abundance of $\psi$ sets a lower limit on $m_\psi$: for $m_\psi \lesssim 150~(120)$ GeV, $g_\psi$ approaches the non-perturbative regime. Indeed, when $m_\psi$ is light, far from the resonance, large values of the coupling $g_\psi$ are required in order to match the correct relic abundance. On the other hand, light DM masses allow for a larger total decay width $\Gamma_{\phi_1}$. Nonetheless, $\Gamma_{\phi_1}$ as large as $\sim 45$--60~GeV turns out to be disfavoured by ID constraints, mainly $\gamma$-ray line searches with the Fermi-LAT satellite. These bounds can be relaxed considering a more conservative DM density distribution thus allowing for a small region of the parameter  space with $m_\psi \sim 200$ GeV where $\Gamma_{\phi_1} \sim 45$ GeV and $\psi$ is under-abundant.
When $m_\psi$ approaches the value $m_{\phi_1}/2$, the annihilation cross section gets enhanced and allowed values of the relic density $\Omega_{\psi} h^2 \lesssim 0.1198$ are achieved with smaller values of $g_\psi$, as low as $\sim 10^{-2} (10^{-3})$ in Scenario 1 (2).

The ID $\gamma$-ray bounds mainly constrain DM masses $m_\psi \lesssim 300$ GeV and  values of the coupling $g_\psi \gsim 0.2$. The bounds from dSphs turn out to be not relevant in this scenarios for perturbative values of $g_\psi$, while searches for  $\gamma$-ray lines give stronger constraints. The relative strength of these bounds is determined by the relative ratios of the effective couplings $c_j^{\phi_1}$. A larger region of the parameter space of both scenarios can be further probed with both $\gamma$-ray lines and dSphs searches by the Fermi-LAT collaboration in the immediate future, by accumulating more data.

\newpage

\section{Conclusions}

A modest excess in the diphoton channel at the invariant mass of about 750 GeV has been reported by both ATLAS and CMS collaborations at the LHC. Motivated by this recent observation, we have considered a model with a heavy parent pseudoscalar decaying into a pair of 750~GeV pseudoscalar resonances.

This hierarchical framework improves the agreement between 8 and 13 TeV data on the resonant production of the 750 GeV (pseudo)scalar.
Moreover, since no additional SM particles seem to accompany to the diphoton signal, we have addressed the possibility for the lighter resonance to decay dominantly into invisible particles, which can play the r\^ole of the DM in the Universe. In this setup, the annihilation of  DM into SM particles proceeds via an $s$-channel exchange of the lighter pseudoscalar. 
We have examined the implications of the diphoton signal on the DM phenomenology, taking into account an array of constraints, both from LHC and from astroparticle physics. 
We have conducted our analysis with an effective theory approach, assuming that the DM is a Majorana fermion and considering two representative scenarios with specific patterns for the effective couplings.

We have fitted the model to the diphoton excess and we have imposed constraints from mono-$X$  ($X$=jet or photon),  dijet and jets plus $E_T^{\mathrm{miss}}$ searches. 
Concerning the DM, we have required compatibility with the relic abundance determined by the PLANCK satellite and with indirect detection constraints from the Fermi-LAT satellite, namely searches for $\gamma$-ray lines from DM annihilation in our Galaxy and for $\gamma$-rays from DM annihilation in dSphs.

We have found that the relic density constraint together with the requirement of perturbativity of the couplings, impose an upper bound on the DM mass $\gtrsim $ 150 GeV (120)  GeV for Scenario 1 (2).
ID bounds further constrain the parameter space, for DM masses  $\lesssim 300$~GeV and values 
of the coupling $g_\psi \gtrsim 0.2$. 
The astroparticle constraints turn out to disfavour a large decay width of the light resonance $\sim$ 45-60 GeV. 

Finally, further constraints imposed using the LHC data and the production of $\phi_1$ and $\phi_2$ provide limits on the coupling of the heavy pseudoscalar to gluons. In Scenario~1 they are placed at 
$c_3^{\phi_2} \lesssim 0.2$ and $\lambda\gtrsim 1000$~GeV. In Scenario~2 on the other  hand, $c_3^{\phi_2} \sim 0.5$--$0.6$ and $\lambda\gtrsim 2000$~GeV. We additionally consider differential distributions of missing transverse energy and transverse momentum of the photon pair. These features can be used to identify the models similar to the ones considered here.

\section*{Acknowledgements} 
VDR is grateful to Giorgio Arcadi for useful discussions.
The authors acknowledge support by the Spanish MINECO through the Centro de Excelencia Severo Ochoa Program under grant SEV-2012-0249, by the Consolider-Ingenio 2010 programme under grant MULTIDARK CSD2009-00064 and by  the Invisibles European ITN project FP7-PEOPLE-2011-ITN, PITN-GA-2011-289442-INVISIBLES.
VDR acknowledges support by the Spanish MINECO  through the project FPA2012-31880 (P.I. Enrique Alvarez Vazquez). VML would like to thank the support by the Spanish MICINN under Grant No. FPA2015-65929-P and the ERC Advanced Grant SPLE under contract ERC-2012-ADG-20120216-320421. J.S.K. has been partially supported by the MINECO (Spain) under contract FPA2013-44773-P and the Consolider-Ingenio CPAN CSD2007-00042. R. RdA is supported by the Ram\'on y Cajal program of the Spanish MICINN and also thanks the support by the ``SOM Sabor y origen de la Materia" (FPA2011-29678), the ``Fenomenologia y Cosmologia de la Fisica mas alla del Modelo Estandar e lmplicaciones Experimentales en la era del LHC" (FPA2010-17747) MEC projects and the Severo Ochoa MINECO project SEV-2014-0398. 

\begin{appendix}
\section{Decay formulae}
For the forthcoming discussion it is convenient to recast the effective couplings of both pseudoscalar states $\phi_{1,2}$ to the SM gauge bosons in Eq.~(\ref{eq:gauge}):
\beq
c_{\gamma \gamma}^i &=& c^{\phi_i}_1 \cos^2 \theta_W + c^{\phi_i}_2 \sin^2 \theta_W \, , \label{eq:coup1}\\ 
c_{ZZ}^i &=& c^{\phi_i}_1 \sin^2 \theta_W + c^{\phi_i}_2 \cos^2 \theta_W \, , \label{eq:coup2}\\
c_{Z\gamma}^i &=& 2 (c^{\phi_i}_2-c^{\phi_i}_1) \sin \theta_W  \cos \theta_W \, , \label{eq:coup3}\\
c_{WW}^i &=& c^{\phi_i}_2\, , \label{eq:coup4}\\
c_{gg}^i &=& c^{\phi_i}_3\,, \label{eq:coup5}
\eeq
where $\sin\theta_W$ denotes the sine of the Weinberg angle.

Given the Lagrangian described in Eq.~(\ref{eq:gauge}) and using Eqs.~(\ref{eq:coup1})--(\ref{eq:coup5}) one can obtain the partial decay widths for the two pseudoscalar particles $\phi_1$ and $\phi_2$ decaying to the final states $i$ and $j$ by the general formula:
\begin{equation}
\Gamma_{\phi_{1,2} \rightarrow i j} = s_{ij} \,  \frac{\squared{\mathcal{M}_{\phi_{1,2} \rightarrow i j}} }{16 \pi \,  m_{\phi_{1,2}}}   \sqrt{1- 2 \frac{(m_i^2 + m_j^2)}{m_{\phi_{1,2}}^2} + \frac{(m_i^2 - m_j^2)^2}{m_{\phi_{1,2}}^4} }  \ ,
\label{eq:dGamma}
\end{equation}
where the statistical factor $s_{ij}$ accounts for identical particles in the final state, $\squared{\mathcal{M}_{\phi_{1,2} \rightarrow i j}}$ is the squared matrix element of the process $\phi_{1,2} \rightarrow i j$, and $m_i (m_j)$ is the mass of the particle $i(j)$. 

The squared matrix elements for the decay processes of $\phi_1$ read:
\begin{align}
\squared{\mathcal{M}_{\phi_1 \; \rightarrow \; g g}} = & \, 256 \frac{c^2_{gg}}{\Lambda_{\phi_1}^2}s^2   \ ,  \\
\squared{\mathcal{M}_{\phi_1 \; \rightarrow \; W^+ W^-}} = & \,  64 \frac{c^2_{WW}\,s^2}{\Lambda_{\phi_1}^2} \left(1 - \frac{4 m_W^2}{s} \right) \ , \\
\squared{\mathcal{M}_{\phi_1 \; \rightarrow \; Z Z}} = & \,  32 \frac{c^2_{ZZ} \, s^2}{\Lambda_{\phi_1}^2} \left(1 - \frac{4 m_Z^2}{s} \right) \ , \\
\squared{\mathcal{M}_{\phi_1 \; \rightarrow \; Z \gamma}}  = & \, 
16 \frac{c^2_{Z \gamma} \, s^2}{\Lambda_{\phi_1}^2} \left( 1 - \frac{m_Z^2}{s} \right)^2 \ , \\
\squared{\mathcal{M}_{\phi_1 \; \rightarrow \; \gamma\gamma}} = & \, 32 \frac{c^2_{\gamma\gamma} \, s^2}{\Lambda_{\phi_1}^2}   \, \\
\squared{\mathcal{M}_{\phi_1 \; \rightarrow \; \psi \psi}} = & 8  \, g_{\psi}^2 \,  s    \ ,
\label{eq:MPchichi} 
\end{align}
while the ones for $\phi_2$ are given by,
\begin{align}
\squared{\mathcal{M}_{\phi_2 \; \rightarrow \; g g}} = & \, 256 \frac{c^2_{gg}}{\Lambda_{\phi_1}^2}s^2   \ ,  \\
\squared{\mathcal{M}_{\phi_2 \; \rightarrow \; W^+ W^-}} = & \,  64 \frac{c^2_{WW}\,s^2}{\Lambda_{\phi_1}^2} \left(1 - \frac{4 m_W^2}{s} \right) \ , \\
\squared{\mathcal{M}_{\phi_2 \; \rightarrow \; Z Z}} = & \,  32 \frac{c^2_{ZZ} \, s^2}{\Lambda_{\phi_1}^2} \left(1 - \frac{4 m_Z^2}{s} \right) \ , \\
\squared{\mathcal{M}_{\phi_2 \; \rightarrow \; Z \gamma}}  = & \, 
16 \frac{c^2_{Z \gamma} \, s^2}{\Lambda_{\phi_1}^2} \left( 1 - \frac{m_Z^2}{s} \right)^2 \ , \\
\squared{\mathcal{M}_{\phi_2 \; \rightarrow \; \gamma\gamma}} = & \, 32 \frac{c^2_{\gamma\gamma} \, s^2}{\Lambda_{\phi_1}^2}   \, \\
\squared{\mathcal{M}_{\phi_2 \; \rightarrow \; \phi_1 \phi_1}} = & 4  \, \lambda^2    \, .
\label{eq:MP2} 
\end{align}
Where $s$ is the center of mass energy that for an on-shell decay of the pseudoscalar particles are $s=m_{\phi_1}^2$ and $s=m_{\phi_2}^2$ respectively. \\

The relevant decay widths for the pseudoscalar $\phi_1$ are obtained using Eq.~(\ref{eq:dGamma}):
\begin{align}
\Gamma_{\phi_1 \; \rightarrow \; g g}= & \, \frac{8c_{gg}^2 m_{\phi_1}^3}{\pi \Lambda_{\phi_1}^2},\\
\Gamma_{\phi_1 \; \rightarrow \; W^+ W^-}=\, & \frac{2c_{WW}^2 m_{\phi_1}^3}{\pi \Lambda_{\phi_1}^2}\left(1-\frac{4m_W^2}{m_{\phi_1}^2}\right)^{3/2},\\
\Gamma_{\phi_1 \; \rightarrow \; Z Z}=\, &\frac{c_{ZZ}^2 m_{\phi_1}^3}{\pi \Lambda_{\phi_1}^2}\left(1-\frac{4m_Z^2}{m_{\phi_1}^2}\right)^{3/2},\\
\Gamma_{\phi_1 \; \rightarrow \; Z \gamma}= &\,\frac{c_{Z\gamma}^2 m_{\phi_1}^3}{2 \pi \Lambda_{\phi_1}^2}\left(1-\frac{m_Z^2}{m_{\phi_1}^2}\right)^{3},\\
\Gamma_{\phi_1 \; \rightarrow \; \gamma\gamma}= &\, \frac{c_{\gamma\gamma}^2 m_{\phi_1}^3}{\pi \Lambda_{\phi_1}^2},\\
\Gamma_{\phi_1 \; \rightarrow \; \psi \psi}= &\, \frac{g_{\psi}^2m_{\phi_1}}{4\pi}\left(1-\frac{4m_\psi^2}{m_{\phi_1}^2}\right)^{1/2}.
\end{align}
   
The decay widths for the heavy pseudoscalar $\phi_2$ are:
\begin{align}
\Gamma_{\phi_2 \; \rightarrow \; g g}= & \, \frac{8c_{gg}^2 m_{\phi_2}^3}{\pi \Lambda_{\phi_2}^2},\\
\Gamma_{\phi_2 \; \rightarrow \; W^+ W^-}=\, & \frac{2c_{WW}^2 m_{\phi_2}^3}{\pi \Lambda_{\phi_2}^2}\left(1-\frac{4m_W^2}{m_{\phi_2}^2}\right)^{3/2},\\
\Gamma_{\phi_2 \; \rightarrow \; Z Z}=\, &\frac{c_{ZZ}^2 m_{\phi_2}^3}{\pi \Lambda_{\phi_2}^2}\left(1-\frac{4m_Z^2}{m_{\phi_2}^2}\right)^{3/2},\\
\Gamma_{\phi_2 \; \rightarrow \; Z \gamma}= &\,\frac{c_{Z\gamma}^2 m_{\phi_2}^3}{2 \pi \Lambda_{\phi_2}^2}\left(1-\frac{m_Z^2}{m_{\phi_2}^2}\right)^{3},\\
\Gamma_{\phi_2 \; \rightarrow \; \gamma\gamma}= &\, \frac{c_{\gamma\gamma}^2 m_{\phi_2}^3}{\pi \Lambda_{\phi_2}^2},\\
\Gamma_{\phi_2 \; \rightarrow \; \phi_1 \phi_1}= &\, \frac{\lambda^2}{8\pi m_{\phi_2}}\left(1-\frac{4m_{\phi_1}^2}{m_{\phi_2}^2}\right)^{1/2}.
\end{align}

\end{appendix}

\end{document}